\documentclass[a4paper,fleqn,usenatbib]{mnras}
\usepackage{graphicx}
\usepackage{amssymb}
\usepackage{amsmath}
\usepackage[T1]{fontenc}
\usepackage{ae,aecompl}
\usepackage{natbib,times}
\usepackage{lscape}
\usepackage{array}
\usepackage{setspace}
\usepackage{helvet}
\usepackage{float}
\newcommand{\PreserveBackslash}[1]{\let\temp=\\#1\let\\=\temp}
\newcolumntype{C}[1]{>{\PreserveBackslash\centering}p{#1}}
\newcolumntype{R}[1]{>{\PreserveBackslash\raggedleft}p{#1}}
\newcolumntype{L}[1]{>{\PreserveBackslash\raggedright}p{#1}}

\newcommand{\mc}{\multicolumn}

\begin{document}

\title[Dynamical state for 964 galaxy clusters]
      {Dynamical state for 964 galaxy clusters from Chandra X-ray images}

\author[Yuan \& Han]
       {Z. S. Yuan$^{1,2}$
    and J. L. Han$^{1,2,3}$ \thanks{E-mail: hjl@nao.cas.cn} 
\\
1. National Astronomical Observatories, Chinese Academy of Sciences, 
20A Datun Road, Chaoyang District, Beijing 100012, China\\
2. CAS Key Laboratory of FAST, NAOC, Chinese Academy of Sciences,
           Beijing 100101, China \\
3. School of Astronomy, University of Chinese Academy of Sciences,
           Beijing 100049, China 
}

\date{Accepted XXX. Received YYY; in original form ZZZ}

\label{firstpage}
\pagerange{\pageref{firstpage}--\pageref{lastpage}}
\maketitle


\begin{abstract}
The dynamical state of galaxy clusters describes if clusters are
relaxed dynamically or in a merging process of subclusters. By using
archival images from the {\it Chandra X-ray Observatory}, we derive a
set of parameters to describe the dynamical state for 964 galaxy
clusters. Three widely used indicators for dynamical state, the
concentration index $c$, the centroid shift $\omega$ and the power
ratio $P_3/P_0$ are calculated in the circular central region with a
radius of 500 kpc. We also derive two adaptive parameters, the profile
parameter $\kappa$ and the asymmetry factor $\alpha$, in the best
fitted elliptical region. The morphology index $\delta$ is then
defined by combining these two adaptive parameters, which indicates
the dynamical state of galaxy clusters and has good correlations to
the concentration index $c$, the centroid shift $\omega$, the power
ratio $P_3/P_0$, and the optical relaxation factor $\Gamma$. For a
large sample of clusters, the dynamical parameters are continuously
distributed from the disturbed to relaxed states with a peak in the
between, rather than the bimodal distribution for the two states. We
find that the newly derived morphology index $\delta$ works for the
similar fundamental plane between the radio power, cluster mass and
the dynamical state for clusters with diffuse radio giant-halos and
mini-halos. The offset between masses estimated from the
Sunyaev-Zel\'dovich effect and X-ray images depends on dynamical
parameters. All dynamical parameters for galaxy clusters derived from
the {\it Chandra} archival images are available on
http://zmtt.bao.ac.cn/galaxy\_clusters/dyXimages/.
\end{abstract}


\section{Introduction}

In the hierarchical structure formation scenario, galaxy clusters are
formed by continuous merging of smaller infalling groups and finally
achieve dynamical equilibrium \citep[e.g.][]{ps74, mbb+09, cog13}.
As the largest gravitational bound systems in the universe, clusters
of galaxies consist of a large number of member galaxies embedded in
the intracluster medium (ICM) and the dark matter halo. 
Often seen inside a galaxy cluster are the substructures in the
spatial distribution of galaxies \citep[e.g.,][]{ds88,wh13}.
Because the hot gas in the ICM observed in the X-ray band is in
kinematics equilibrium under the assumption of the energy density
equipartition between the ICM and galaxies, many substructures have
also been detected in the X-ray images  \citep[e.g.,][]{kbp+01}.

The dynamical state of galaxy clusters describes if clusters are
relaxed dynamically or are still in a merging process of subclusters.
Clusters of galaxies are therefore broadly divided into relaxed
clusters and dynamically disturbed clusters. Relaxed clusters roughly
reach the virial equilibrium, showing a symmetrical structure in the
distribution of galaxies and the intracluster gas around their
brightest cluster galaxy (BCG).
While dynamically disturbed clusters, such as the Bullet cluster
\citep{mgd+02}, obviously deviate from the virial equilibrium with
subclusters around two or more very bright galaxies
\citep[e.g.,][]{cd96,bbr+02,wh13} or clear substructures in X-ray
images \citep[e.g.,][]{mefg95,mjfv08,yda+16}.
The dynamical state of galaxy clusters is ideally expressed by the
three-dimensional (3D) velocity distribution of member galaxies
\citep[e.g.][]{cd96,ets+10,evn+12} or the ICM
\citep[e.g.][]{db06,lyt+15,lyt+16,yda+16}. In practice, the dynamical
state can be roughly indicated by the 1D radial velocity or redshift
distribution
\citep[e.g.,][]{yv77,ds88,wb90,ssg99,hmp+04,hph+09,rph18,yds+18}, or
the 2D positions of member galaxies in the sky plane projected from
their real spatial distribution
\citep[e.g.,][]{gb82,wb90,fk06,rbp+07,as10,evn+12,wh13,ltl+18} or the
projected hot gas distribution shown in X-ray or microwave-band images
of clusters \citep[e.g.,][]{mefg95,kbp+01,bfs+05,cbs12,cds+18}.

Substructures in such a projected two dimensional distribution of
member galaxies or hot gas have been parameterized quantitatively to
describe the dynamical state of galaxy clusters, such as the asymmetric
factors or the probability distribution based on the optical data
\citep[e.g.][]{wb90,rbp+07,wh13}, or the concentration index
\citep[e.g.,][]{mefg95,srt+08}, the centroid shift \citep[e.g.,][]{
  mefg95,pfb+06,mjfv08} and the power ratio \citep[e.g.,][]{bt95,
  bt96, bpa+10} based on the data of the X-ray images, see details in
Section~\ref{knownpara}.

\citet{wh13} derived relaxation parameters for 2092 rich clusters
based on photometric data by using the asymmetry, the ridge flatness
and the normalized deviation of the smoothed optical maps, which is
the largest sample of clusters to our knowledge with dynamical
parameters.
\citet{kgm+19} found a good agreement between the galaxy density maps
and the X-ray brightness maps, and therefore estimated the dynamical
parameters for 890 clusters by the method given by \citet{wh13}.
Recently, \citet{cds+18} also tried to parameterize the morphological
characterization of synthetic maps of the Sunyaev-Zel\'dovich (SZ)
effect for a sample of 258 simulated clusters.
\citet{ltl+18} estimated optical substructure parameters for 72
clusters which have {\it Chandra} X-ray images and/or {\it Planck}
Sunyaev-Zel\'dovich maps by using optical imaging and spectroscopic
data.
\citet{rlt+18} showed the close correlation between dynamical
parameters (difference between the optical and X-ray center, the
projected offset of the BCG from other cluster centroids) and the
offset from the scaling relations for clusters.
\citet{zhk+20} also calculated the offsets between BCGs and gas center
as traced by SZ effect ({\it SPT}) and/or X-ray ({\it Chandra} and
{\it XMM-Newton}) images as dynamical parameter for 288 massive
clusters.

The dynamical parameters have been involved to many studies, e.g. the
cosmological constraints based on galaxy clusters with different X-ray
morphologies \citep[e.g.][]{mefg95},
the evolution of galaxy clusters in the cosmological model of
structure formation \citep[e.g.][]{bt96},
the presence of radio halos and relics in merging clusters and the
formation of mini-halos in relaxed clusters
\citep[e.g.,][]{ceg+10,ceb+13,ccb+15, yhw15},
the mass estimation \citep[e.g.,][]{rlt11} and radial mass profile
\citep[e.g.,][]{bap+19} for clusters of galaxies,
the scaling relations for cluster mass estimations
\citep[e.g.,][]{omb+06,crb+07,zfb+08,zjc+13},
the dynamical-state dependence of galaxy luminosity functions
\citep[e.g.,][]{wh15},
and activity of super massive black holes in BCGs
\citep{kvc+15,yhw16}.

In fact, any definition of dynamical parameters may be biased because
we cannot get the real 3D distribution and velocities. Nevertheless,
the identification and analysis of substructures of any projected
galaxy distribution or X-ray images should give the lower limit of
dynamical parameters ideally derived from the unavailable real 3D
distribution.  Therefore, it is still valuable to get dynamical
parameters from the X-ray images of galaxy clusters.
The {\it Chandra} satellite has observed about 1000 clusters of
galaxies in the X-ray band, and the high resolution {\it Chandra}
image is ideally to reveal the substructures from the ICM.
In this paper, we derive a set of dynamical parameters for 964
clusters based on the archival data of the {\it Chandra X-ray
  Observatory}.

\begin{figure}
\centering
\includegraphics[angle=270,width=0.42\textwidth]{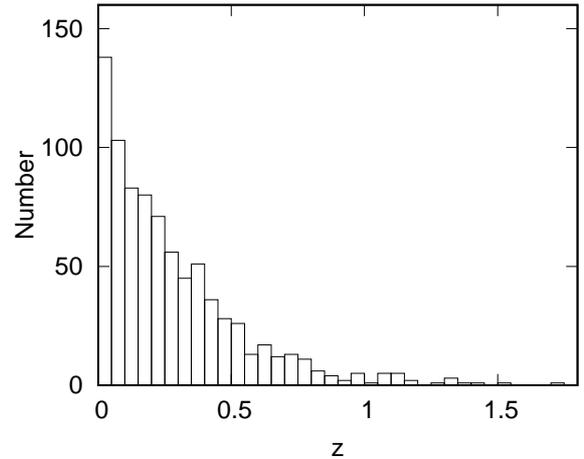}
\caption{Redshift distribution for 911 of 964 clusters with
    available redshift.}
\label{redshift}
\end{figure}

In Section 2, we describe how we get and process the {\it Chandra}
data. In Section 3, we define and calculate various dynamical
parameters. Comparison of the derived parameters and discussions on
the impact of dynamical state on the scaling relations are presented
in Section 4. A summary is given in Section 5. Throughout this paper,
we assume a flat $\Lambda$CDM cosmology taking $H_0=70$ km~s$^{-1}$
Mpc$^{-1}$, $\Omega_m=0.3$ and $\Omega_{\Lambda}=0.7$.

\section{Data collection and processing}

\subsection{Chandra data for galaxy clusters}
\label{datacoll}
We collect the observation information for clusters of galaxies
archived by the {\it Chandra} in two approaches. 

First, we get the information for all observations which categorized
as ``Clusters of galaxies'' in the {\it Chandra} archival
database\footnote{http://cda.harvard.edu/chaser/}. As a result, more
than 2000 observations are obtained, often several observations are
made for one cluster. For clusters observed more than one time, we
generally choose the observation with largest photon number. However,
we select the observation for nearby clusters (mostly with $z<0.05$)
not only with a sufficient photon number but also with a good coverage
of CCDs. After removing reduplicate observations, we obtained data for
915 clusters.

Second, we cross match several widely used cluster catalogs: the Abell
\citep{aco89}, MCXC \citep{pap+11}, redMaPPer \citep{rrb+14}, CAMIRA
\citep{o14}, WH15 \citep{whl12,wh15} and WHY18 \citep{why18}, with
archived {\it Chandra} observations to find serendipitous observed
clusters with criteria as following: (1) the search radius from the
nominal center of a {\it Chandra} observation is set as 10 arcminutes
so that a cluster in the field of view of a {\it Chandra} ACIS-I
observation with a side-length of 16.9 arcminutes can be covered, (2)
the exposure time is larger than 10 kiloseconds and the average count
rate is larger than 2 Hz, so that there are enough photons to work on,
(3) the data status should be archived, (4) observations in all
science categories of {\it Chandra} are selected except for ``Clusters
of galaxies'' (the category has been done), (5) the exposure
instrument is limited with ACIS and without any gratings, (6) the
observation types are constrained as GO (Guest Observer) and GTO
(Guaranteed Time Observation), and the exposure mode is set as TE
(Timed Exposure). With these steps, after merging the outputs for
different cluster catalogs and checking the actual coordinates in
X-ray images, we finally obtain another 49 serendipitously observed
clusters.

\label{imgproc}
\begin{figure*}
\centering
\includegraphics[angle=270,width=55mm]{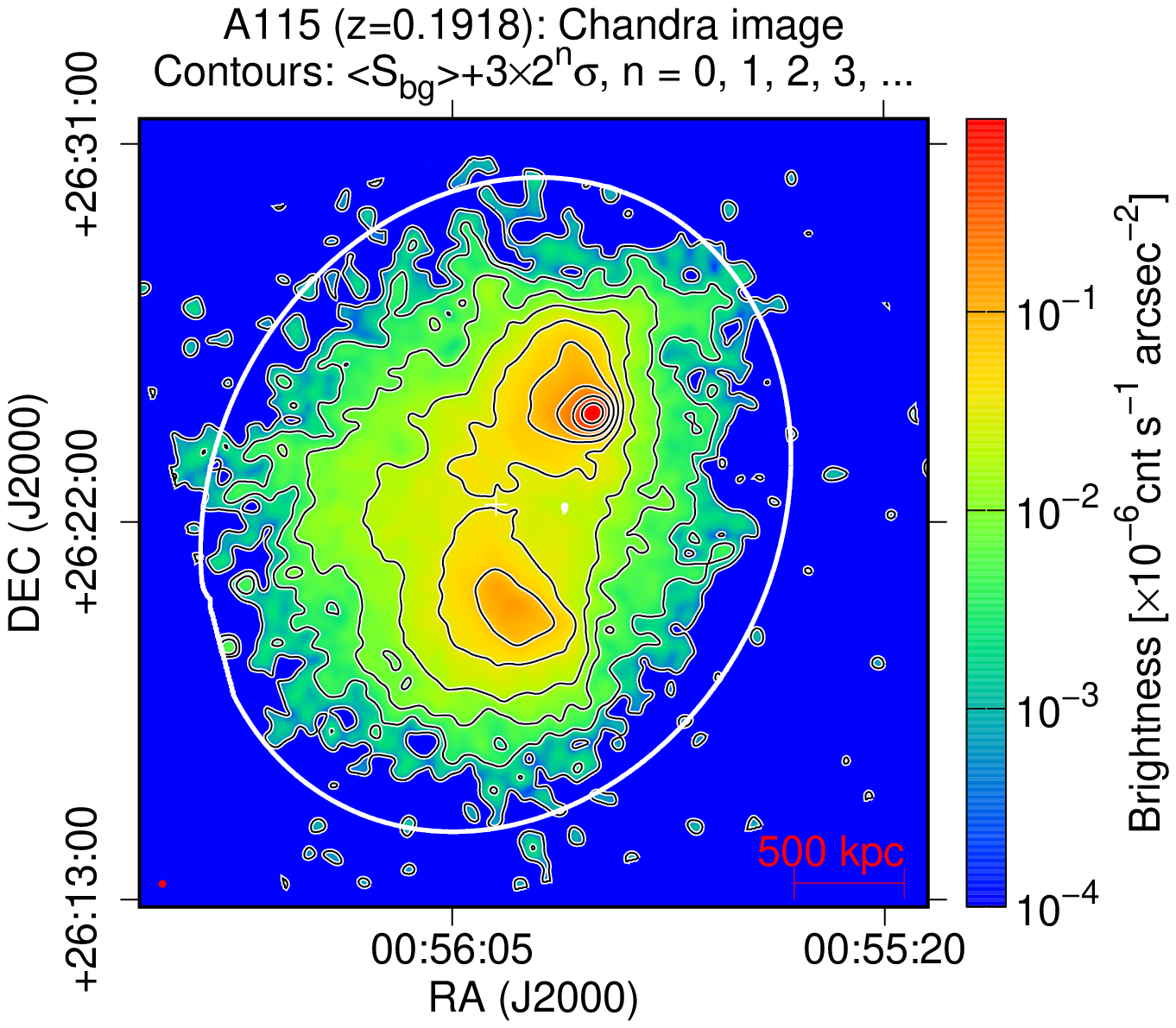}
\includegraphics[angle=270,width=55mm]{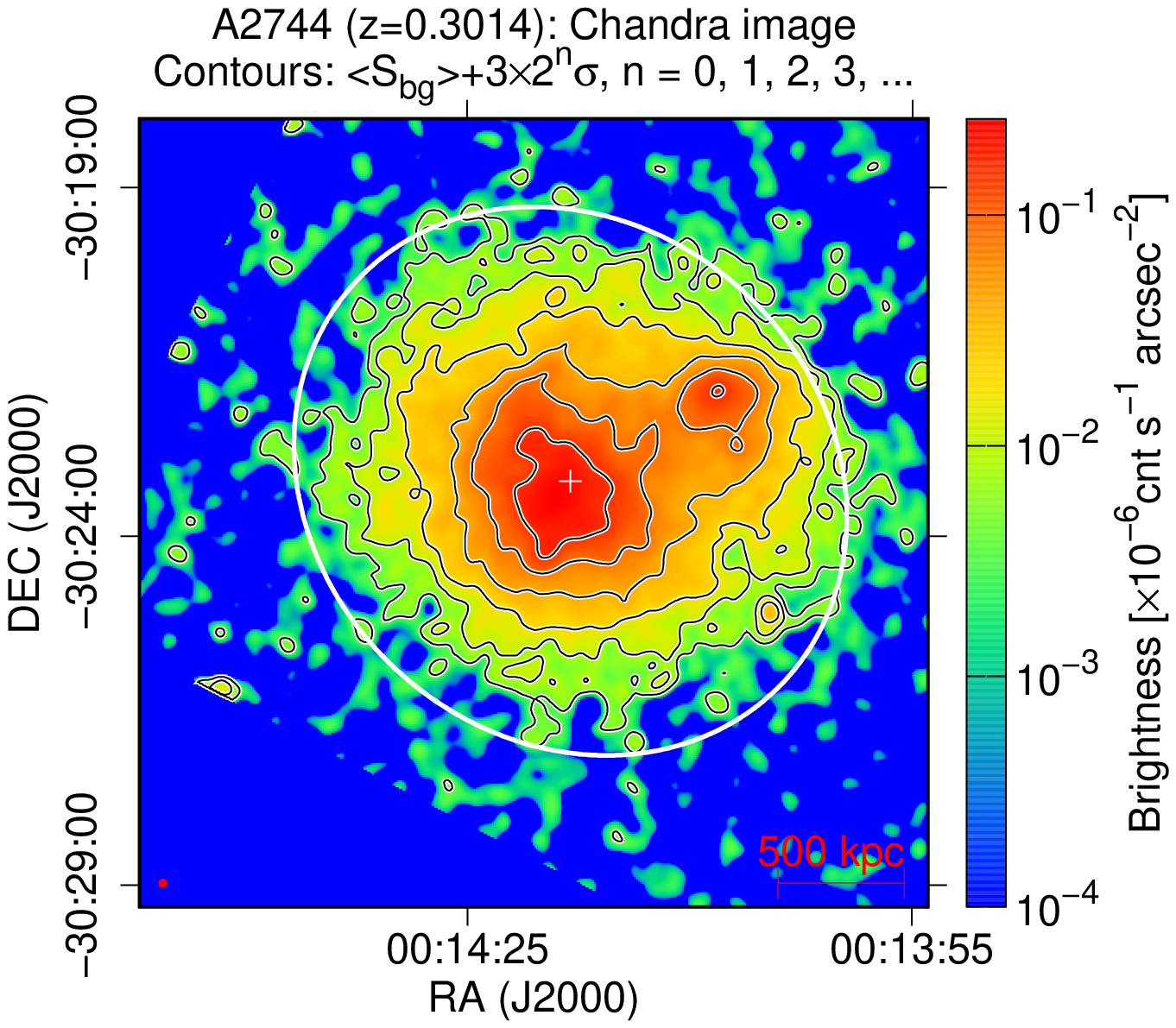}  
\includegraphics[angle=270,width=55mm]{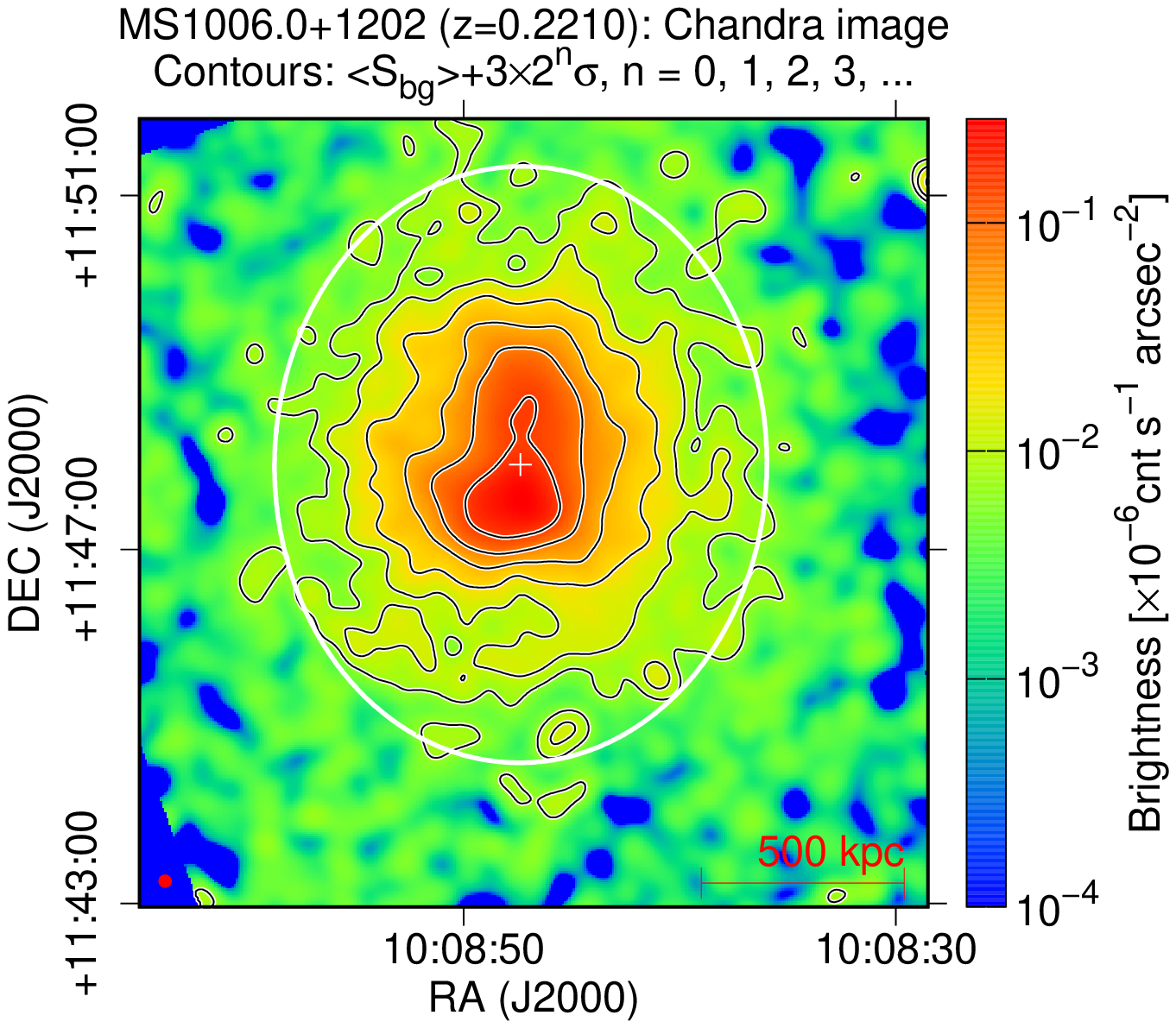}\\[2mm]  
\includegraphics[angle=270,width=55mm]{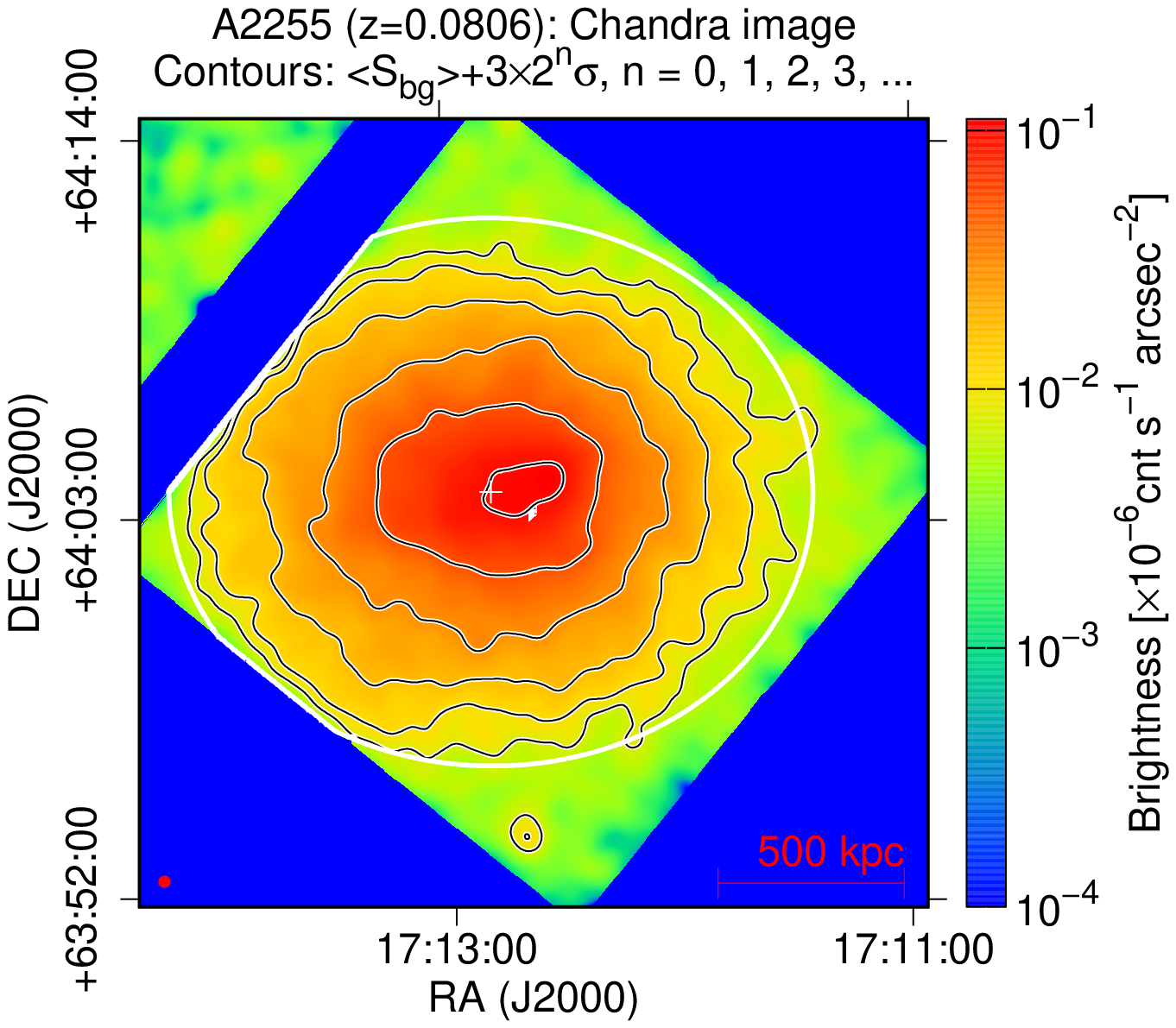} 
\includegraphics[angle=270,width=55mm]{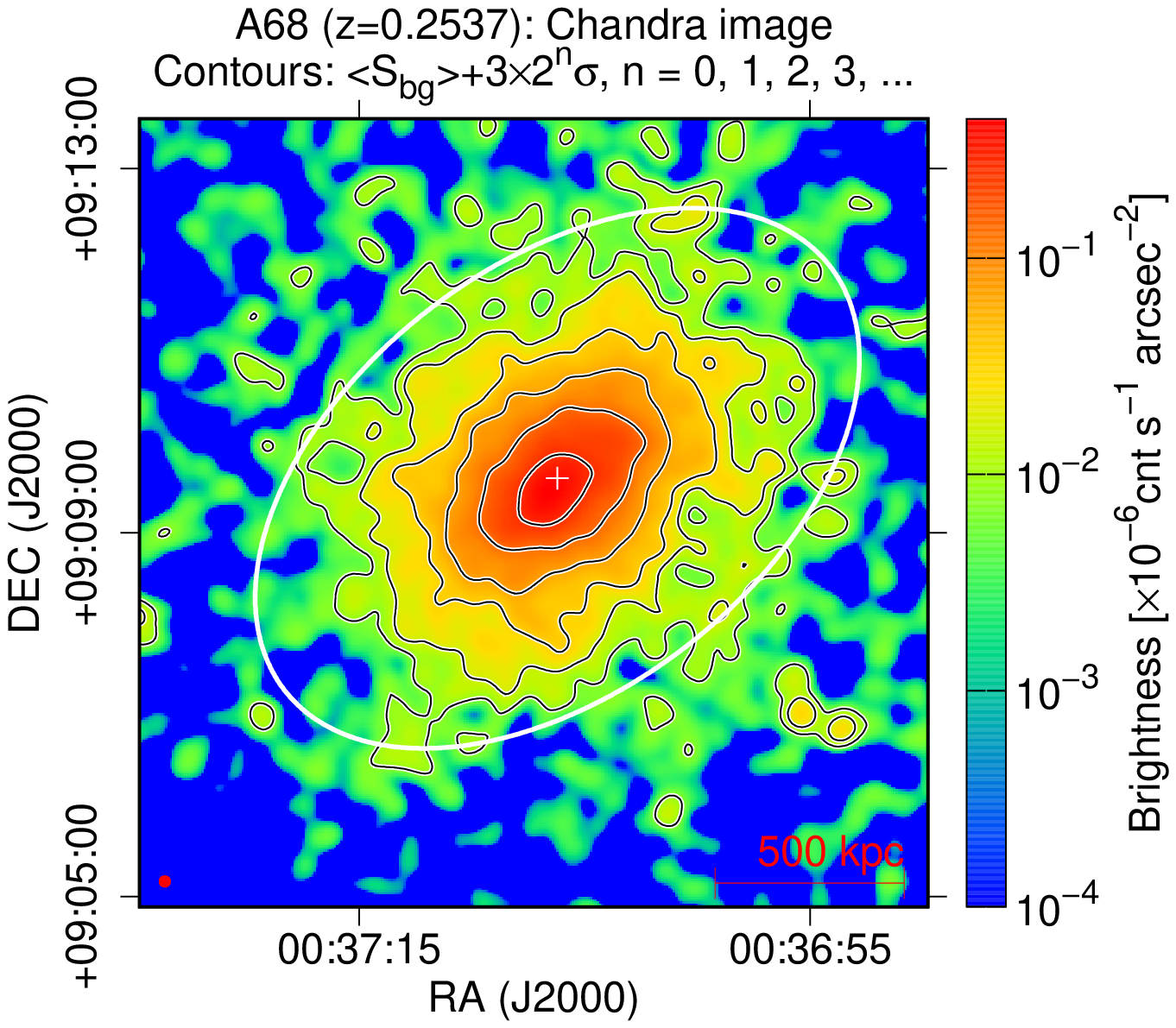} 
\includegraphics[angle=270,width=55mm]{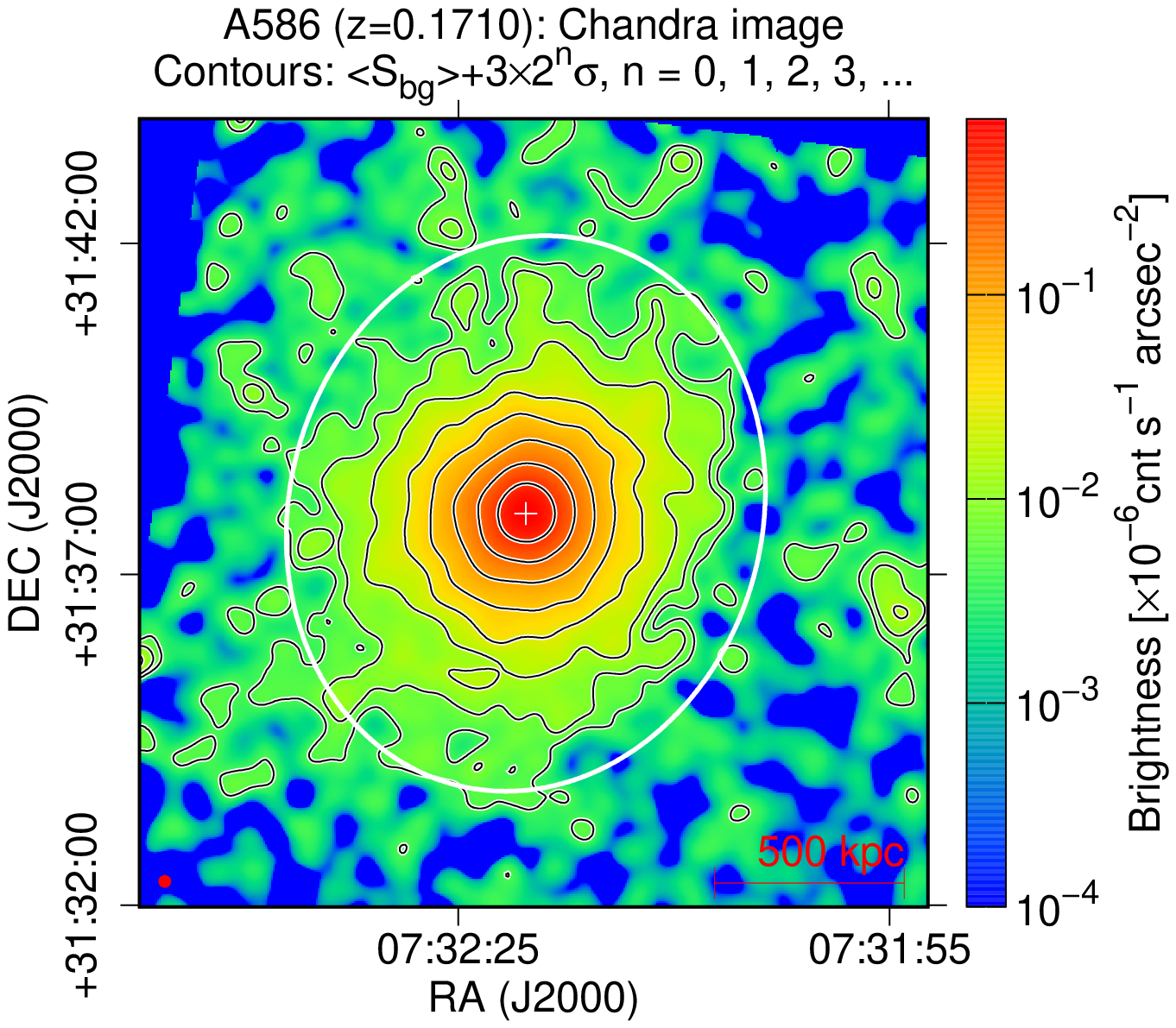}\\[2mm]  
\caption{Example images for 6 galaxy clusters, and images for all
  clusters are available on the web page
  http://zmtt.bao.ac.cn/galaxy\_clusters/dyXimages/. Cluster name,
  redshift and contour levels are written on the top of each
  panel. Surface brightness is logarithmically indicated as the color
  bar on the right of each panel. We select a clean region around
  every clusters to calculate the mean brightness $\langle S_{bg}
  \rangle$, and its fluctuations $\sigma$. The white cross stands for
  the center $(x_0,y_0)$ of the best fitted model. The adaptive
  dynamical parameters are calculated within the white ellipse, which
  also defines the actual region for the morphology parameters of
  clusters. The red circle in the bottom-left corner of each panel
  indicates the smooth scale generally with a diameter of 30 kpc. The
  scale of 500 kpc is plotted on the bottom-right corner to indicate
  the physical scale of clusters, which is also the radius size used
  to calculate the concentration index, the centroid shift and the
  power ratio.}
\label{examples}
\end{figure*}

In total, we have a sample of 964 galaxy clusters with good {\it
  Chandra} X-ray image data (see Table 1), which mostly have a lower
redshift less than 0.5 (see Figure~\ref{redshift}) but some high
redshift clusters up to $z\sim1.5$ (e.g., SPT-CL J2040-4451) do come
out. The estimated masses of some clusters \citep[e.g.,][]{pap+11} are
in the range from $10^{13} M_{\odot}$ (e.g., MCXC J1242.8+0241) to
$2.0\times10^{15} M_{\odot}$ (e.g., MACS J0417.5-1154).

\begin{table*}
\setlength{\tabcolsep}{1mm}
\caption{Dynamical parameters for 964 clusters of galaxies (see
  http://zmtt.bao.ac.cn/galaxy\_clusters/dyXimages/ for the full table).}
\footnotesize
\begin{center}
  \begin{tabular}{lrrrrcccccr}
  \hline
  \mc{1}{l}{Name}  &\mc{1}{c}{obsID} &\mc{1}{c}{RA} &\mc{1}{c}{DEC} &\mc{1}{c}{$z$}  &\mc{1}{c}{${\rm log}_{10}(c)$} &\mc{1}{c}{${\rm log}_{10}(\omega)$}  &\mc{1}{c}{${\rm log}_{10}(P_3/P_0)$} &\mc{1}{c}{$\kappa$} &\mc{1}{c}{${\rm log}_{10}(\alpha)$} &\mc{1}{c}{$\delta$}\\
  \mc{1}{l}{(1)} &\mc{1}{c}{(2)} &\mc{1}{c}{(3)} &\mc{1}{c}{(4)} &\mc{1}{c}{(5)} &\mc{1}{c}{(6)} &\mc{1}{c}{(7)} &\mc{1}{c}{(8)} &\mc{1}{c}{(9)} &\mc{1}{c}{(10)} &\mc{1}{c}{(11)}\\
  \hline
  SPT-CLJ0000-5748          &18238   &0.25000 &-57.80695 &0.7020 &-0.29$\pm$0.01 &-3.50$\pm$0.01 &-8.15$\pm$0.22  &  0.80 & -1.19$\pm$0.01 &-0.02$\pm$0.01\\
  SPT-CLJ0001-5440          &19761   &0.40583 &-54.66972 &0.7300 &-0.50$\pm$0.01 &-1.53$\pm$0.01 &-3.99$\pm$0.05  &  1.40 & -0.34$\pm$0.01 & 1.00$\pm$0.01\\
  A2717                     & 6974   &0.80042 &-35.92722 &0.0490 &-0.30$\pm$0.01 &-2.91$\pm$0.01 &-7.58$\pm$0.01  &  1.40 & -1.94$\pm$0.01 &-0.09$\pm$0.01\\
  Z15                       &12251   &1.58453 & 10.86429 &0.1663 &-0.53$\pm$0.01 &-2.51$\pm$0.01 &-7.14$\pm$0.02  &  1.15 & -1.76$\pm$0.01 &-0.15$\pm$0.01\\
  ACT-CLJ0008.1+0201        &19586   &2.04333 &  2.02000 &0.3651 &-0.67$\pm$0.01 &-2.44$\pm$0.01 &-6.46$\pm$0.07  &  2.00 & -0.88$\pm$0.01 & 1.07$\pm$0.01\\
  WHLJ001037.1+112957       &20514   &2.65464 & 11.49916 &0.1042 &-0.62$\pm$0.01 &-3.16$\pm$0.01 &-4.82$\pm$0.01  &  1.91 & -0.70$\pm$0.01 & 1.13$\pm$0.01\\
  A2734                     & 5797   &2.83625 &-28.85500 &0.0625 &-0.60$\pm$0.01 &-2.53$\pm$0.01 &-6.80$\pm$0.01  &  1.49 & -1.50$\pm$0.01 & 0.28$\pm$0.01\\
  ZwCl0008.8+5215           &19916   &2.85667 & 52.52806 &0.1040 &-1.07$\pm$0.01 &-1.04$\pm$0.01 &-5.42$\pm$0.01  &  1.44 & -0.81$\pm$0.01 & 0.71$\pm$0.01\\
  MACS-J0011.7-1523         & 6105   &2.92875 &-15.38944 &0.3780 &-0.41$\pm$0.01 &-3.05$\pm$0.01 &-7.11$\pm$0.05  &  0.96 & -1.72$\pm$0.01 &-0.26$\pm$0.01\\
  A7                        &15157   &2.93861 & 32.41566 &0.1026 &-0.70$\pm$0.01 &-2.47$\pm$0.02 &-6.77$\pm$0.03  &  1.80 & -1.63$\pm$0.01 & 0.42$\pm$0.01\\
\hline
\end{tabular}
\label{tab1}
\end{center}
{Notes. Columns: (1) cluster name; (2) observation ID of Chandra;
  (3-4) right ascension and declination in J2000; (5) redshift; (6)
  the concentration index; (7) the centroid shift; (8) the power
  ratio; (9) the profile parameter; (10) the asymmetry factor; (11)
  the morphology index.}
\end{table*}

\subsection{Image processing}

The {\it Chandra} data for these clusters of galaxies are processed in
the standard manner by using CIAO 4.9 \citep{fma+06} and calibration
database CALDB 4.7.2. Data files of all observations are downloaded
with CIAO command {\tt download\_chandra\_obsid}, and reprocessed by
{\tt chandra\_repro} to make sure the latest calibration products are
used. To be consistent with previous works \citep[e.g.,][]{srt+08},
the photons are filtered in 0.5-5 keV band. To remove flare events, we
use the tool {\tt lc\_clean} to cut out data in time intervals where
the photon count rate deviates from their mean value more than
20\%. For few clusters with giant flares, the mean count rate of
photons is not a good reference, thus time intervals for giant flares
are removed manually. Point sources are detected automatically with
the CIAO routine {\tt wavdetect} with a proper threshold which should
not be too high to miss some faint point sources, or not too low to
lead a false detection of some fluctuations as being point
sources. The size of point source model should not be too small to
miss some relatively large member or foreground galaxies, or too large
to confuse the core of relaxed clusters as point sources. We check the
list of point sources very carefully and then remove the real point
sources. We use the tool {\tt dmfilth} to fill ``holes'' of subtracted
point sources according to the brightness of ambient regions. Images
of clusters are also exposure corrected and background subtracted
accordingly. Since the pixel stands for different physical scales for
different redshifts, the X-ray images of clusters are smoothed to 30
kpc by a Gaussian function, except for 4 very close clusters NGC 1395,
Virgo, MCXC J1242.8+0241 and MCXC J1315.3-1623 with $z<0.01$ for which
the smooth scale are set to 10 kpc. For few clusters without available
redshifts, the images are smoothed to 5 arcseconds. The examples of
final X-ray images are shown in Figure~\ref{examples}.

\section{Dynamical parameters from X-ray images}
\label{dp}

In this section, we obtain three widely used dynamical parameters,
i.e., the concentration index $c$, the centroid shift $\omega$, the
power ratio $P_3/P_0$, and compare our results with those available in
literature. Then, we define two adaptive dynamical parameters, i.e.,
the profile parameter $\kappa$ and the asymmetry factor $\alpha$, in
the best fitted elliptical region. We combine them to make the
morphology index $\delta$, which is an excellent indicator for the
dynamical state of galaxy clusters.

\subsection{Three widely used dynamical parameters}
\label{knownpara}

\subsubsection{The concentration index, $c$}
\label{conindex}
Relaxed clusters usually host a very luminous cool core in their
center \citep[e.g.,][]{fnc84,f94,mbb+12}, while the core of disturbed
clusters generally have been destroyed by violent merger
events. \citet{srt+08} defined the concentration index as the ratio of
X-ray fluxes integrated in the centroid and the whole regions of
galaxy clusters to quantify their dynamical state. The concentration
index $c$ is calculated in two circular regions with the core radius
of 100 kpc and the outer radius of 500 kpc
\citep[e.g.,][]{ceg+10,ceb+13}, i.e.,
\begin{equation}
  c=\frac{S_{100~\rm kpc}}{S_{500~\rm
      kpc}}=\frac{\sum\limits_{R<100\rm~kpc}f_{\rm
      obs}(x_i,y_i)}{\sum\limits_{R<500\rm~kpc}f_{\rm obs}(x_i,y_i)},
\label{c}
\end{equation}
where $f_{\rm obs}(x_i,y_i)$ means the observed X-ray flux at pixel
$(x_i,y_i)$. With the definition in Equation~\ref{c}, we set the two
radii as being 100 kpc and 500 kpc, and calculate the concentration
index $c$ for clusters in our sample, as listed in Table 1. The
concentration index for some low-redshift clusters (e.g., the Vigro
cluster) cannot be calculated due to the size of 500 kpc is much out
of the coverage of the {\it Chandra} CCD, and they are denoted as
``--'' in the full Table~1.

\begin{figure}
  \begin{center}
\includegraphics[angle=0,width=0.23\textwidth]{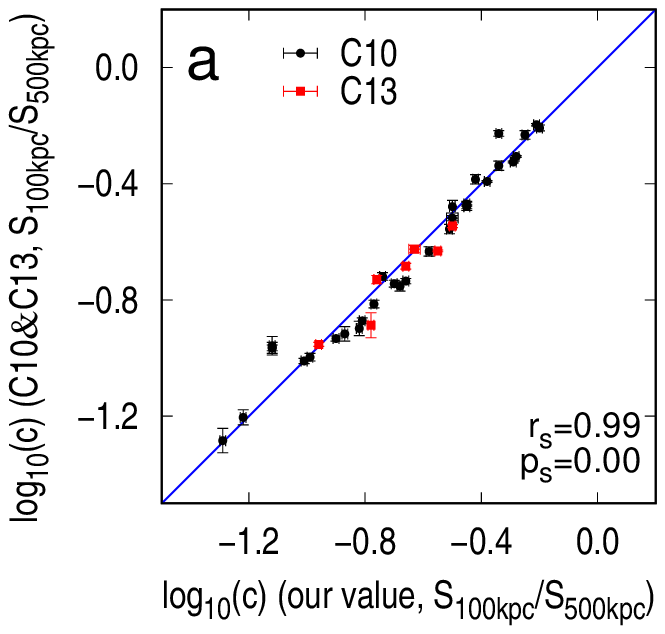}
\includegraphics[angle=0,width=0.23\textwidth]{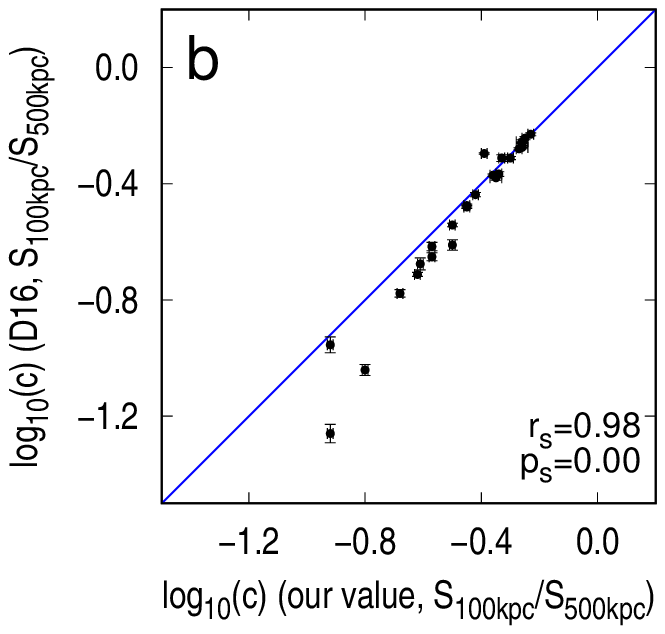}\\[2mm]
\includegraphics[angle=0,width=0.23\textwidth]{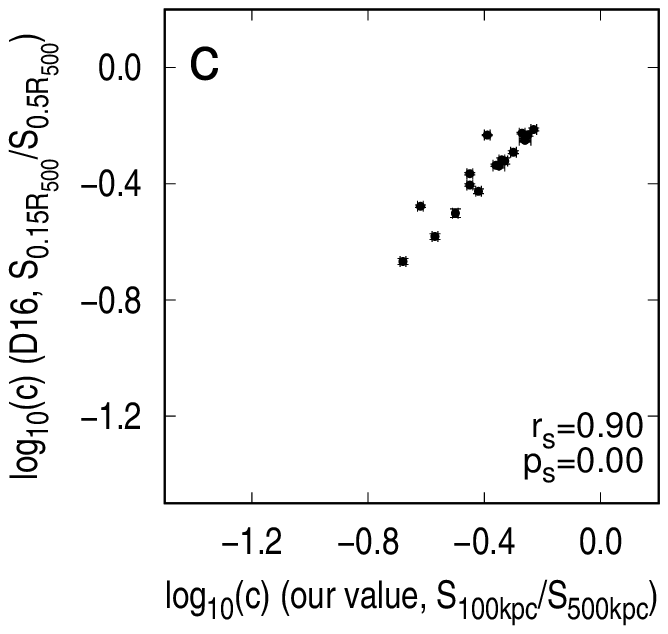}
\includegraphics[angle=0,width=0.23\textwidth]{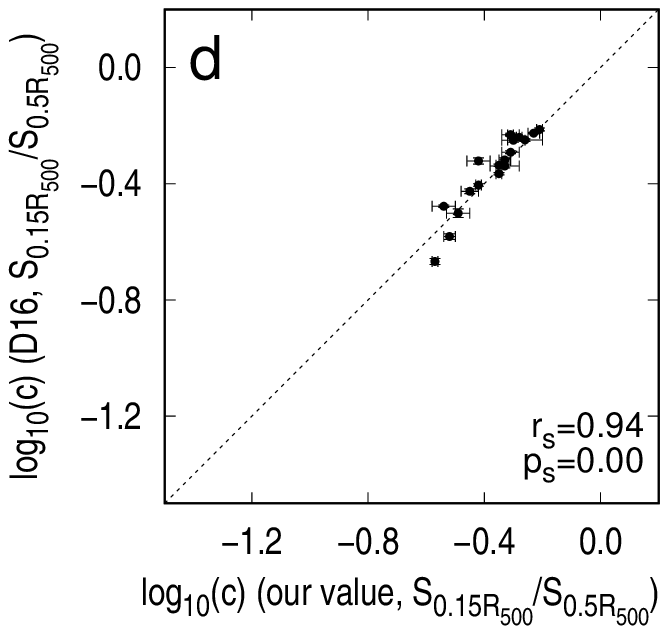}\\[2mm]
\includegraphics[angle=0,width=0.23\textwidth]{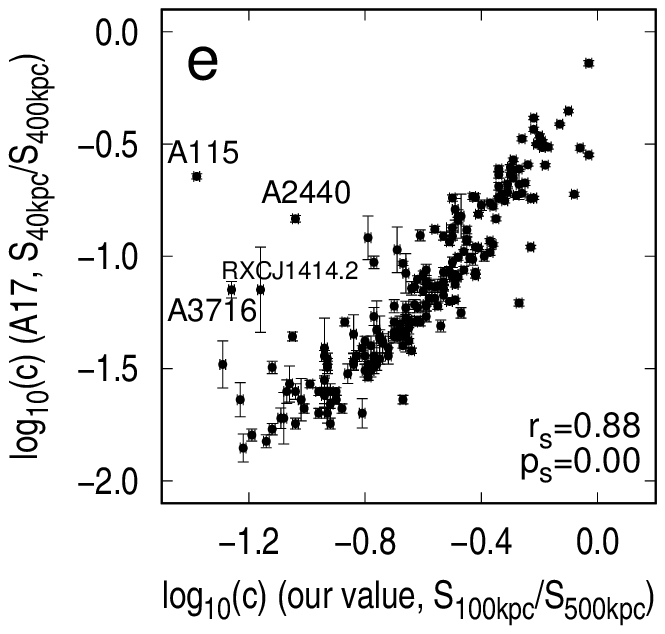}
\includegraphics[angle=0,width=0.23\textwidth]{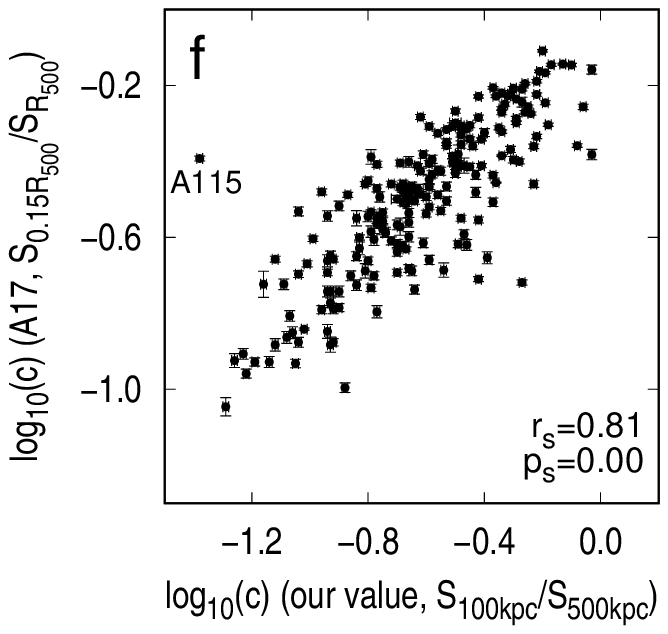}\\[2mm]
\includegraphics[angle=0,width=0.23\textwidth]{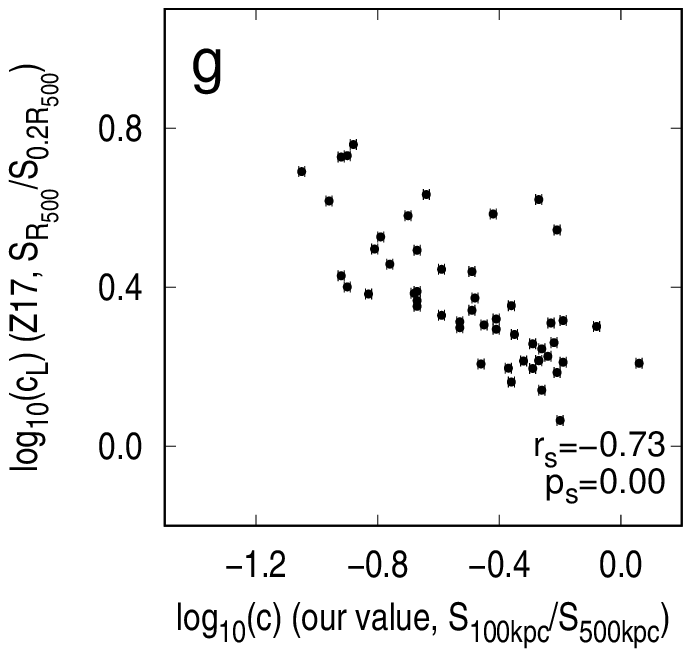}
\includegraphics[angle=0,width=0.23\textwidth]{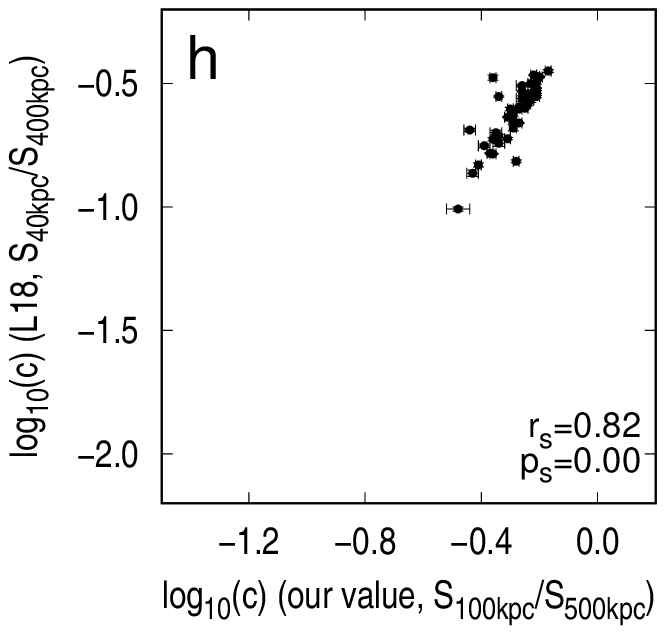}
\caption{Comparison for the concentration indexes that we obtained
  with those in literatures. The solid line in panels a and b and
  dotted line in panel d indicate equivalent values in X and Y axes
  because the values in both axes are calculated with the same radii
  for the center and outer regions. The Spearman rank-order
  correlation coefficient $r_{\rm s}$ and the relevant significance
  $p_{\rm s}$ are labelled in the right-bottom corner of each
  panel. Data for the X-axis are our values and for the Y-axis are
  obtained from: C10=\citet{ceg+10}, C13=\citet{ceb+13},
  D16=\citet{der+16}, A17=\citet{ajf+17}, Z17=\citet{zrs+17} and
  L18=\citet{lty+18}.}
\label{ccorr}
\end{center}
\end{figure}

Although the concentration index has been derived by a lot of authors
from {\it Chandra} and/or {\it XMM-Newton} images, the radii of the
central and the outer regions are chosen differently. In
Figure~\ref{ccorr}, we compare our values with those calculated in
literatures. \citet{ceg+10} defined the concentration index as
$c=S_{100\rm~kpc}/S_{500\rm~kpc}$ and calculated the $c$ for 32
clusters based on the {\it Chandra} images and for 7 clusters in a
later paper \citep{ceb+13}. Since we set the same core and outer radii
as \citet{ceg+10,ceb+13}, we obtain almost the same values (see
Figure~\ref{ccorr}a). Very good correlations can be found between our
values of $c=S_{100\rm~kpc}/S_{500\rm~kpc}$ and those obtained by
\citet{der+16} in Figure~\ref{ccorr}b for 25 clusters who also defined
the core radius of 100 kpc and the outer radius of 500 kpc, and in
Figure~\ref{ccorr}c and Figure~\ref{ccorr}d for 19 clusters with the
core radius of $0.15~R_{500}$ and the outer radius of $0.5~R_{500}$
which we also calculated. Here $R_{500}$ is the radius of a galaxy
cluster within which the matter density of a cluster is 500 times of
the critical density of the universe. \citet{ajf+17} worked on a large
sample of 214 clusters with the {\it Chandra} data, and calculated the
concentration index with two definitions, i.e.,
$c=S_{40\rm~kpc}/S_{400\rm~kpc}$ and
$c=S_{0.15~R_{500}}/S_{R_{500}}$. Again, good consistences are found
between our values of $c=S_{100\rm~kpc}/S_{500\rm~kpc}$ and those in
\citet{ajf+17}, see Figure~\ref{ccorr}e and Figure~\ref{ccorr}f. The
outliers, i.e., A115, A2440, A3716 and RXC J1414.2+7115, in
Figure~\ref{ccorr}e and Figure~\ref{ccorr}f are merging clusters with
a bi-model brightness distribution (see the image of A115 in
Figure~\ref{examples} as an example). For these clusters we take the
best-fitted center (see Section 3.2.1) located between the two
subclusters similar to \citet{ceg+10}, while \citet{ajf+17} took the
X-ray peak of one subcluster as the center, which cause the value
difference. There are 53 clusters in common among our sample and
\citet{zrs+17}, inverse correlation appears with large scatter in
Figure~\ref{ccorr}g because they define reversely as $c_{\rm
  L}=S_{R_{500}}/S_{0.2~R_{500}}$. \citet{lty+18} calculated
$c=S_{40\rm~kpc}/S_{400\rm~kpc}$ for 41 clusters from the {\it
  Chandra} images. We also get a good agreement between our values and
results of \citet{lty+18} for 41 clusters as shown in
Figure~\ref{ccorr}h. The Spearman rank-order correlation coefficient
$r_{\rm s}$, and the relevant significance $p_{\rm s}$, are marked
inside each panel to show the reliability of correlations
\citep[see][p. 640]{ptvf92}. Here the small value of $p_{\rm s}$
indicates a significant correlation between the two values.

The concentration index can be calculated within the widely used,
mass-related radius $R_{500}$, but there are some limitations for a
large sample of clusters. First, the radius $R_{500}$ is the byproduct
during the mass estimation for clusters, which (1) requires redshifts
that for 53 clusters in our sample are missing, (2) needs good quality
of X-ray image with sufficient photons for spectrum analysis and (3)
is estimated with the assumption of spherically symmetry and virial
equilibrium that is invalid for many clusters as seen in this
paper. Second, the aperture with a radius of $R_{500}$ is also too
large for the {\it Chandra} CCD coverage for many clusters. The median
$R_{500}$ for 1,742 clusters in \citet{pap+11} is about 0.8 Mpc, the
corresponding aperture exceeds 10 arcminutes when $z<0.15$ (note that
the scale of the whole field of view for ACIS-I is 16.9'$\times$16.9'
and ACIS-S is 8.3'$\times$50.6').

\begin{figure}
\begin{center}
\includegraphics[angle=0,width=0.23\textwidth]{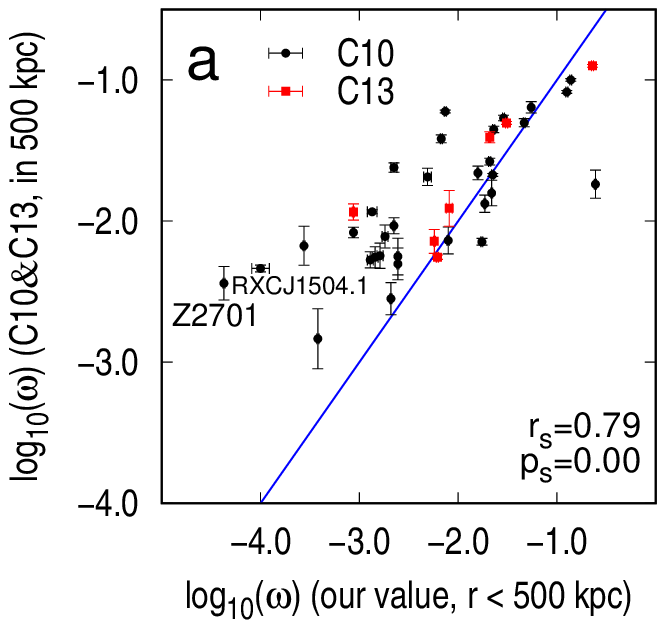}
\includegraphics[angle=0,width=0.23\textwidth]{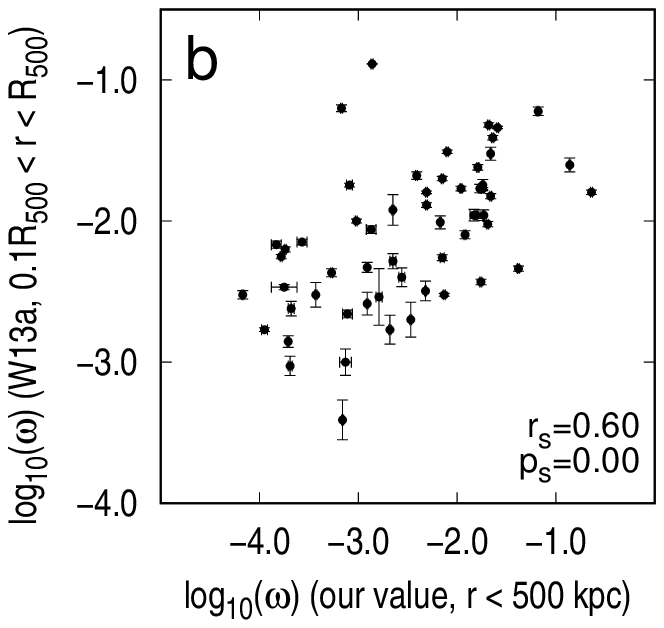}\\[2mm]
\includegraphics[angle=0,width=0.23\textwidth]{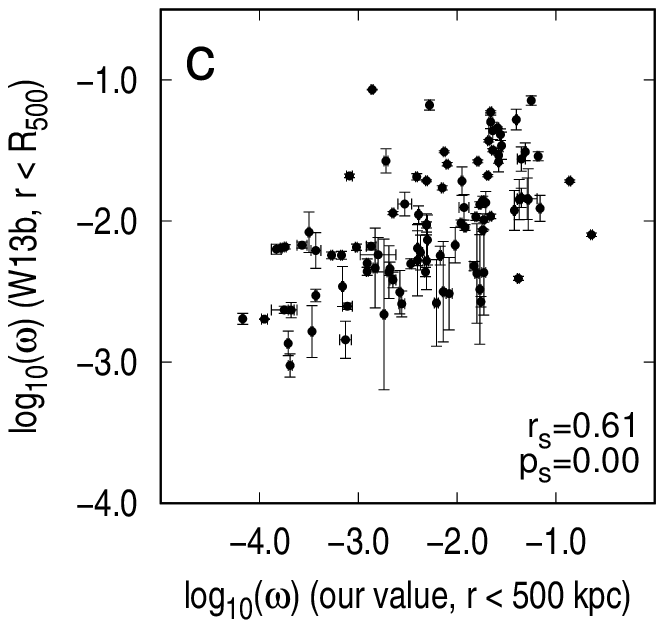}
\includegraphics[angle=0,width=0.23\textwidth]{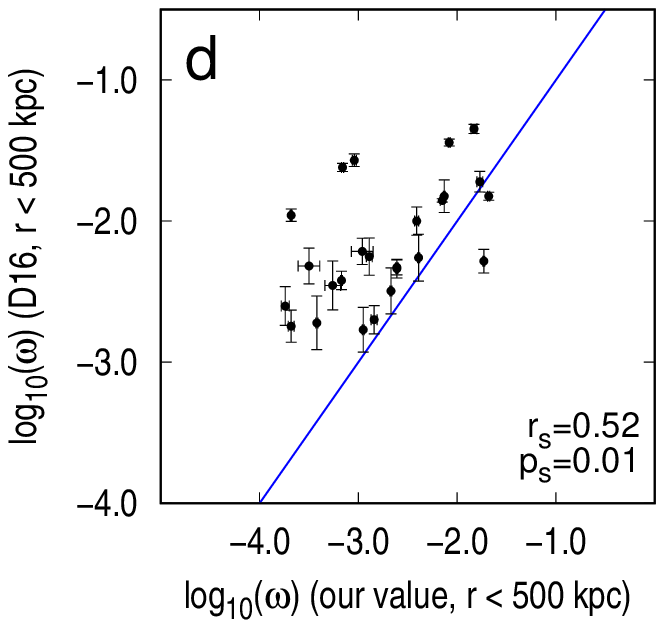}\\[2mm]
\includegraphics[angle=0,width=0.23\textwidth]{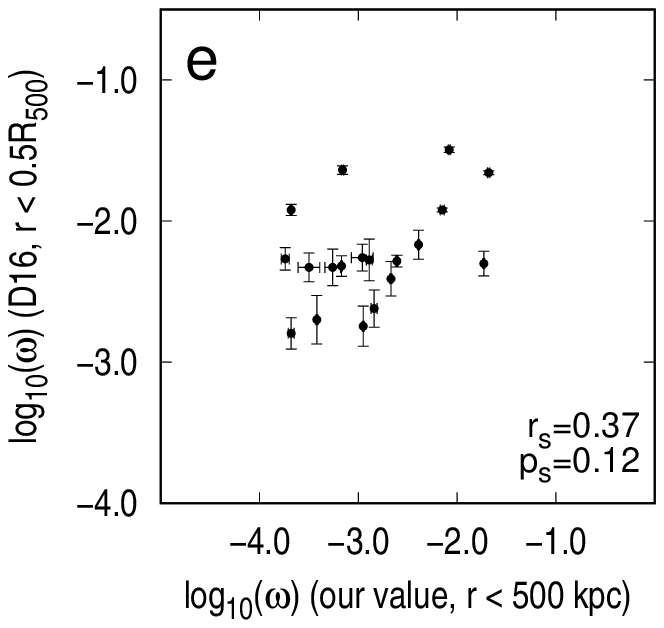}
\includegraphics[angle=0,width=0.23\textwidth]{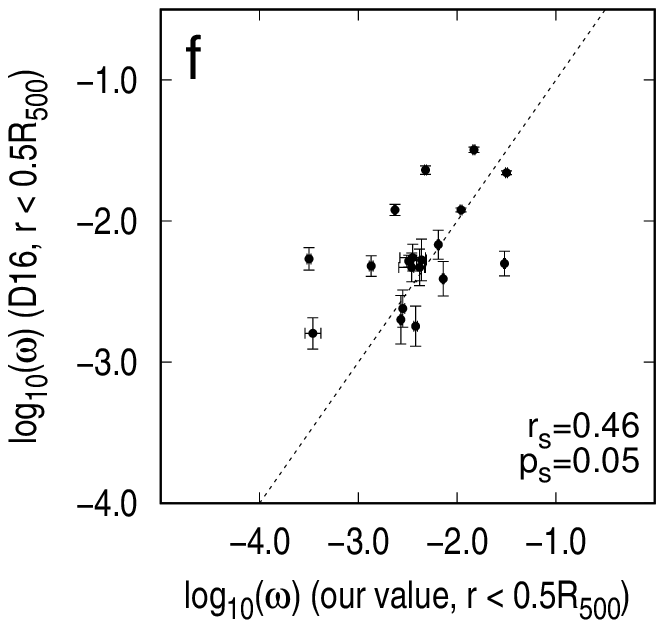}
\caption{Comparison for the centroid shifts we calculated with those in
  literatures. The solid line in the panels a and d indicates
  equivalent values in X and Y axes if the radius is set in the same
  value of 500 kpc, the dotted line in panel f is also an equivalent
  line but for the radius of $0.5~R_{500}$. The correlation parameter
  $r_{\rm s}$ and $p_{\rm s}$ are labelled in the right-bottom corner
  of each panel. Data in the X-axis are our values and in the Y-axis
  are obtained from: C10= \citet{ceg+10}, C13=\citet{ceb+13},
  W13a=\citet{wbsa13}, W13b=\citet{wbc13} and D16=\citet{der+16}.}
\label{wcorr}
\end{center}
\end{figure}

\subsubsection{The centroid shift, $\omega$}
\label{censhift}
The observed X-ray image peak of merging clusters can be off from
their fitted center significantly \citep[e.g.,][]{mfg93,kbp+01}, while
the deviation is usually inconspicuous for relaxed
clusters. \citet{pfb+06} defined the centroid shift $\omega$ as the
standard deviation of the projected separation between the X-ray
brightness peak and the model fitted center, which is computed in a
series of circular aperture centered on the X-ray brightness peak from
$0.05~R_{\rm ap}$ to $R_{\rm ap}$ in steps of $0.05~R_{\rm ap}$
\citep[see also][]{omb+06}, thus
\begin{equation}
\omega=[\frac{1}{n-1}\sum\limits_{i}(\Delta_{i}-\langle \Delta
  \rangle)^2]^{\frac{1}{2}}\times \frac{1}{R_{\rm ap}}.
\label{w}
\end{equation}
Here $R_{\rm ap}=500~\rm{kpc}$ and $n=20$, $\Delta_{i}$ is the
distance between the X-ray brightness peak and the fitted model center
of the $i$th aperture, $\langle\Delta\rangle$ is the mean value of all
$\Delta_{i}$ \citep[e.g.,][]{ceg+10,ceb+13}. We calculate in this work
the centroid shift for clusters in central region with radius of 500
kpc, see Equation~\ref{w} and values are listed in Table 1.

Previously, the centroid shift of clusters has been calculated by
several authors with the similar formula, but for different
regions. In Figure~\ref{wcorr}, the values of centroid shift $\omega$
that we obtained are compared with those calculated in literatures.
\citet{ceg+10} derived the centroid shift for 32 clusters in the
aperture with a radius of 500 kpc, and 7 new clusters later in
\citet{ceb+13}. Good correlation is found between our values and
parameters in \citet{ceg+10,ceb+13}, as shown in Figure~\ref{wcorr}a.
\citet{wbsa13} calculated the centroid shift for 80 clusters by using
the {\it XMM-Newton} images in the annulus region of 0.1--1
$R_{500}$. \citet{wbc13} worked on a larger sample of clusters in
$R_{500}$ which contains 126 clusters based on {\it Chandra} and {\it
  XMM-Newton} data. We get good correlations, with large scatter,
between our values and parameters for 57 clusters in \citet{wbsa13} as
shown in Figure~\ref{wcorr}b, and for 101 clusters in \citet{wbc13} in
Figure~\ref{wcorr}c. \citet{der+16} calculated the centroid shift for
25 clusters in 500 kpc, and for 19 clusters in 0.5 $R_{500}$ with the
{\it Chandra} images. We find clear correlations between our values
and those in \citet{der+16}, as shown in Figure~\ref{wcorr}d and
Figure~\ref{wcorr}e, with large uncertainties. We also calculate the
centroid shift for 19 clusters in \citet{der+16} within the radius of
$0.5~R_{500}$, and compare our values to their results in
Figure~\ref{wcorr}f, and find the data are around the equivalent line
with a better correlation.

In Figure~\ref{wcorr}a and \ref{wcorr}d, the centroid shift are
calculated within the same radius of 500 kpc, but our results seem to
be systematically smaller than those obtained by \citet{ceg+10,ceb+13}
and \citet{der+16}, especially for relaxed clusters. This is mainly
caused by the smoothed images we use. For example, Z2701 and RXC
J1504.1-0248 (marked in Figure~\ref{wcorr}a) have the largest
deviations to the equivalent line. The logarithmic value of the
centroid shift for Z2701 is equal to -4.37 when the image is smoothed
to 30 kpc, but it reduces to -3.47 when the image is smoothed to 10
kpc, and to -2.50 for unsmoothed image which matches the value of
-2.44 obtained by \citet{ceg+10} very well. The logarithm of the
centroid shift for RXC J1504.1-0248 is equal to -4.00 when the image
is smoothed to 30 kpc, but it is reduced to -3.31 when the image is
smoothed to 10 kpc, and to -2.06 for unsmoothed image which is
slightly smaller than the -2.33 obtained by \citet{ceg+10}. The smooth
scale can therefore affect the value of centroid shift
significantly. Considering that the pixel size stands for different
physical scales for clusters with different redshifts, we suggest that
the X-ray images of clusters should be smoothed to a certain physical
size such as the 30 kpc we used.

\begin{figure}
\begin{center}
\includegraphics[angle=0,width=0.23\textwidth]{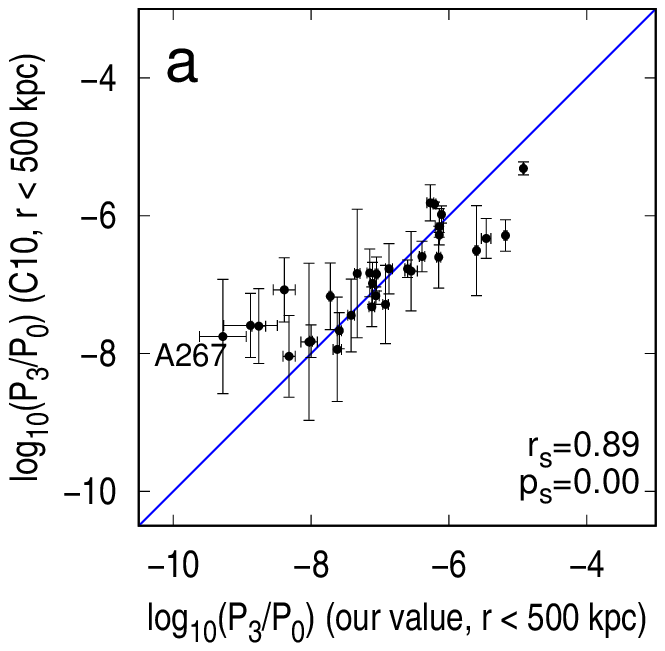}
\includegraphics[angle=0,width=0.23\textwidth]{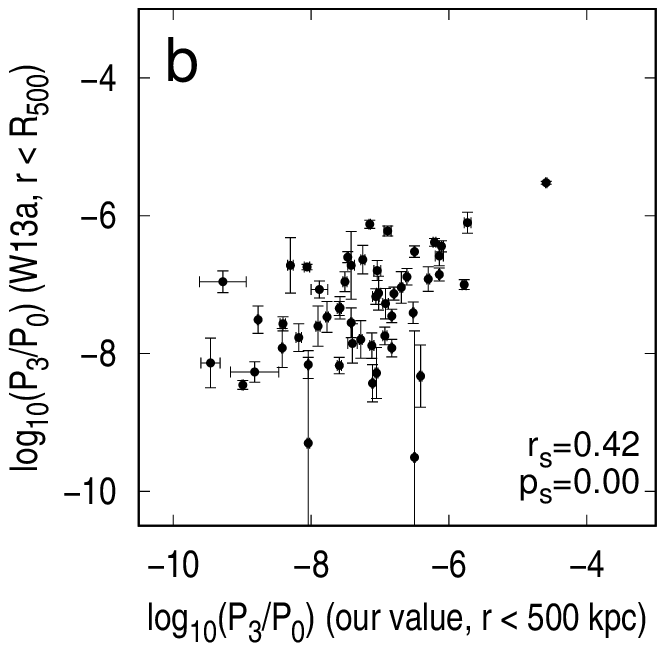}\\[2mm]
\includegraphics[angle=0,width=0.23\textwidth]{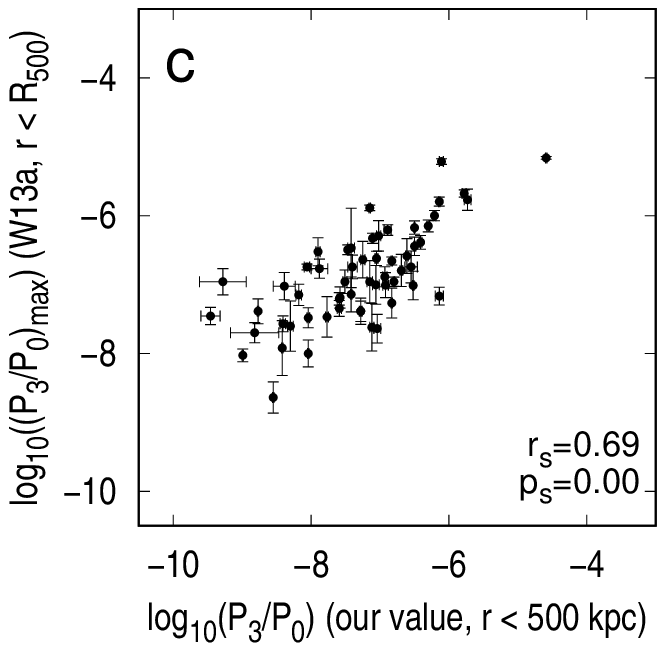}
\includegraphics[angle=0,width=0.23\textwidth]{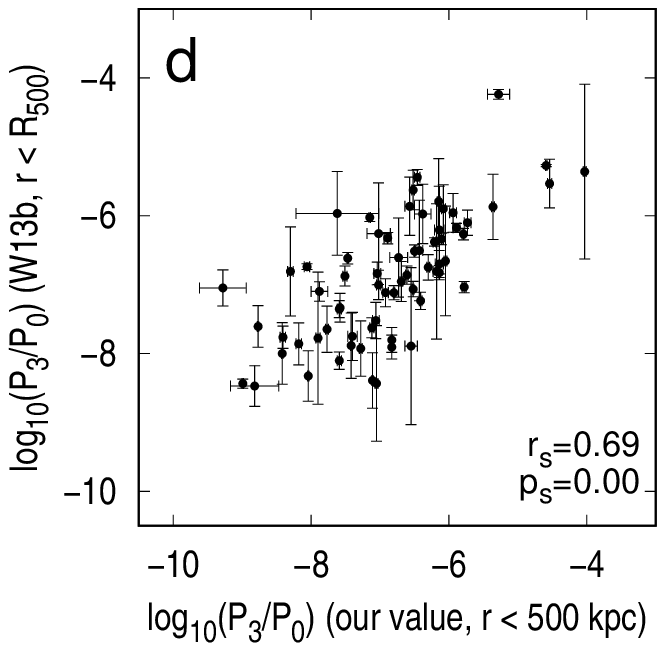}\\[2mm]
\includegraphics[angle=0,width=0.23\textwidth]{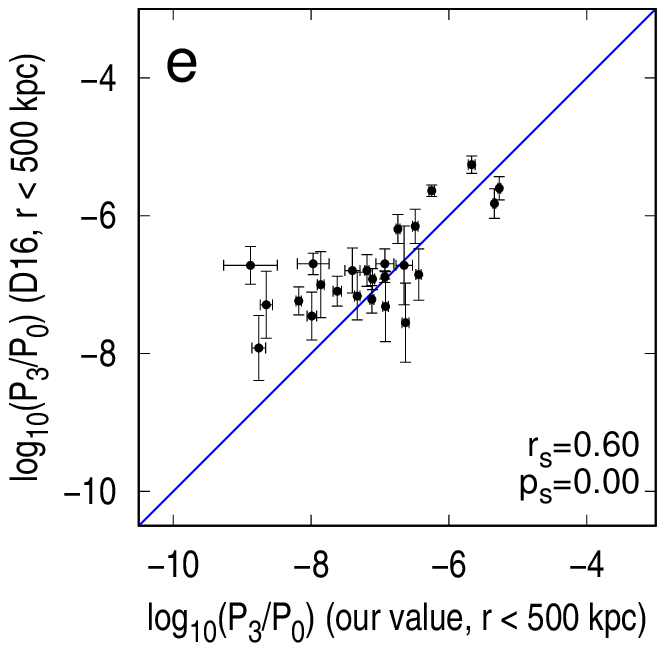}
\includegraphics[angle=0,width=0.23\textwidth]{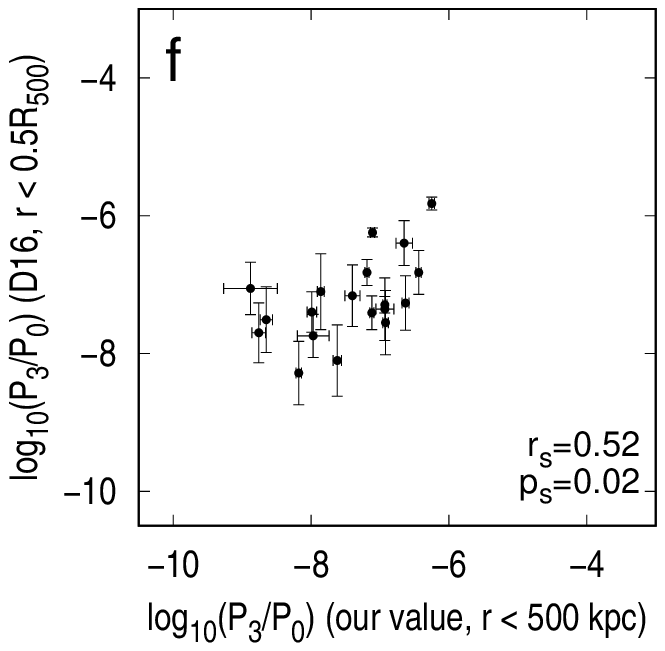}\\[2mm]
\includegraphics[angle=0,width=0.23\textwidth]{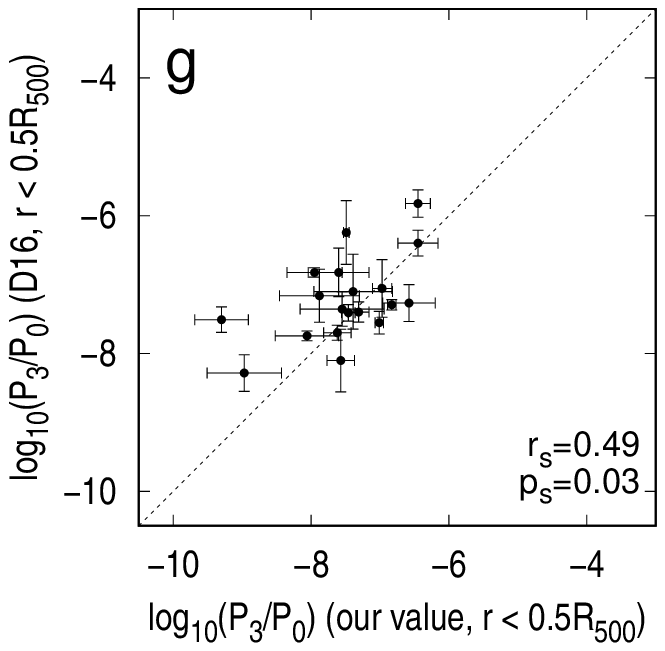}
\caption{Comparison for the power ratios that we calculated with those
  in literatures. The solid line in the panels a and e indicates
  equivalent values in X and Y axes if the radius is set in the same
  value of 500 kpc, the dotted line in panel g is also an equivalent
  line but for the radius of $0.5~R_{500}$. The correlation parameter
  $r_{\rm s}$ and $p_{\rm s}$ are labelled in the right-bottom corner
  of each panel. Data in the X-axis are our values and in the Y-axis
  are taken from: C10=\citet{ceg+10}, W13a=\citet{wbsa13},
  W13b=\citet{wbc13} and D16=\citet{der+16}.}
\label{pcorr}
\end{center}
\end{figure}

\subsubsection{Power ratio, $P_3/P_0$}
\label{powratio}
Because disturbed clusters generally have more remarkable fluctuations
of surface brightness than relaxed clusters, \citet{bt95} defined the
power ratio as dimensionless morphological parameters from the
two-dimensional multipole expansion of the projected gravitational
potential of clusters within $R_{\rm ap}=500\rm kpc$. The moments,
$P_{m}$, are defined as follows:
\begin{equation}
P_{0}=[a_{0}\ln (R_{\rm ap})]^2,
\label{p0}
\end{equation}
\begin{equation}
P_{m}=\frac{1}{2m^{2}R_{\rm ap}^{2m}}(a_{m}^2+b_{m}^2).
\label{pm}
\end{equation}
The moments $a_{m}$ and $b_{m}$ are calculated using
\begin{equation}
a_{m}=\int_{r\le R_{\rm ap}}f_{\rm obs}(x_i,y_i)(r)^{m}\cos(m\theta)dx_idy_i,
\label{am}
\end{equation}
and
\begin{equation}
b_{m}=\int_{r\le R_{\rm ap}}f_{\rm obs}(x_i,y_i)(r)^{m}\sin(m\theta)dx_idy_i,
\label{bm}
\end{equation}
where $f_{\rm obs}(x_i,y_i)$ and $\theta$ have the same meaning as
before.  $P_{3}/P_{0}$ is the power ratio, which was found to be
related to substructures
\citep[e.g.,][]{bfs+05,bpa+10,ceg+10,lfj+17,cds+18}.  Following the
custom, we take the radius at 500 kpc to calculate the power ratio,
see Equation~\ref{p0}-\ref{bm}, and list the results in Table 1.

The values of power ratio that we obtained are compared with those
calculated in literatures in Figure~\ref{pcorr}. Our values show good
consistence to those obtained by \citet{ceg+10} for 32 clusters in
central region with radius of 500 kpc, as shown in
Figure~\ref{pcorr}a. \citet{wbsa13} worked out power ratios for 80
clusters by using the {\it XMM-Newton} images in $R_{500}$, and also
defined $(P_3/P_0)_{\rm max}$ as the maximum value in different
annuluses along the radial direction \citep[see details
  in][]{wbsa13}. We get 57 clusters in common and find good
correlations between our results and those in \citet{wbsa13}, see
Figure~\ref{pcorr}b for $P_3/P_0$ and Figure~\ref{pcorr}c for
$(P_3/P_0)_{\rm max}$. By combining the {\it XMM-Newton} and {\it
  Chandra} data, \citet{wbc13} calculated power ratios for 126 galaxy
clusters in $R_{500}$. The Figure~\ref{pcorr}d shows a clear
correlation between values calculated by us and those in \citet{wbc13}
for 101 clusters in common. \citet{der+16} derived power ratios for 25
clusters in 500 kpc and 19 clusters in $0.5~R_{500}$ from the {\it
  Chandra} images. Good consistencies can be found between our results
and values for clusters from \citet{der+16} in Figure~\ref{pcorr}e and
Figure~\ref{pcorr}f. In Figure~\ref{pcorr}g, we show the correlation
between power ratios in $0.5~R_{500}$ calculated by us and those
obtained by \citet{der+16}.

In Figure~\ref{pcorr}a, \ref{pcorr}e and \ref{pcorr}g, the power
ratios in X-axis and Y-axis are calculated in the same radius, so that
the values obtained by us match well with those calculated by
\citet{ceg+10} and \citet{der+16} around the equivalent line. However,
we get slightly smaller power ratios for relaxed clusters than those
in \citet{ceg+10} and \citet{der+16} because we use the smoothed
images. For example, the logarithm of the power ratio of A267, which
has the largest deviation and labelled in Figure~\ref{pcorr}a, is
equal to -9.28$\pm$0.34 when the image is smoothed to 30 kpc, and it
is reduced to -7.24$\pm$0.48 when we use the unsmoothed image which
matches very well with the value of 7.75$\pm$0.83 obtained by
\citet{ceg+10}.

\subsection{A new morphological parameter for estimating the dynamical state}
In this subsection, we introduce and calculate a new morphological
parameter for estimating the dynamical state of galaxy cluster.

As seen in section \ref{knownpara}, the concentration index $c$, the
centroid shift $\omega$ and the power ratio $P_3/P_0$, which have been
widely used, are mostly calculated in the region of a fixed radius of
500 kpc or $R_{500}$ in recent papers. Obviously the fixed size to 500
kpc is not the best choice because clusters have various
sizes. Second, these three dynamical parameters are defined and
calculated in a circular region but the morphologies for most clusters
are elliptical or even irregular. Third, the redshift information is
needed but may not be available to define the 500 kpc or $R_{500}$,
especially for clusters identified in X-ray band or through SZ
effect. For clusters containing two or more subclusters, the central
region for calculating the dynamical parameters can only focus on the
main subcluster, which is not a proper indicator for the dynamical
state of the whole cluster (such as A115 in Figure~\ref{examples}).

The dynamical state of clusters should be derived in regions adaptive
to their size. The concerned region should be more reasonably taken as
being an elliptical region rather than a circular region. The image
edge of clusters in practice often is shown by the $\langle
S_{bg}\rangle+3\sigma$ contours, as the white ellipse in Figure
\ref{examples}. The dynamical state of clusters should be independent
to the exposure time of X-ray images. In the following, we use two
parameters to describe the global morphological properties of clusters
which are less influenced by the quality of X-ray images. We carry a
few steps toward the goal as following.

\subsubsection{Model fitting for the cluster shape}
\label{modelfit}
To describe the global morphology of a galaxy cluster, we fit the
X-ray surface brightness distribution with an elliptical 2-dimensional
$\beta$-model \citep{cf76,gad+14}:
\begin{equation}
f_{\rm mod}(x_i,y_i)=f_{\rm mod}(r)=A(1+(\frac{r}{r_0})^2)^{-\beta}+C,
\label{Ir}
\end{equation}
where
\begin{equation}
r(x_i,y_i)=\frac{\sqrt{x^2(1-\epsilon)^2+y^2}}{1-\epsilon},
\label{r}
\end{equation}
and
\begin{equation}
  \begin{split}
    x=(x_i-x_0)\cos\theta+(y_i-y_0)\sin\theta,\\
    y=(y_i-y_0)\cos\theta-(x_i-x_0)\sin\theta.
    \label{xy}
\end{split}
\end{equation}
Here $(x_0,y_0)$ are the coordinates of the center of the model, $A$
is the model amplitude, $r_0$ means the core radius, $\beta$ is the
power law index, and $C$ is a constant adjusting the count number for
the average background. $\epsilon$ is the ellipticity of cluster,
$\theta$ is the position angle of cluster, defined as the direction of
major axis from north to east. All these model parameters can be
determined by the least $\chi^2$ fitting.

We fitted the model to the X-ray image of 964 clusters of galaxies,
and obtained the model parameters which define the shape of galaxy
clusters. Among them, the most important is the index $\beta$, which
define steepness of the brightness profile. Others are not very
important for the dynamical state. After the model fitting, the
residual maps show more clearly the disturbed gas distribution. We
have tried some quantitative description of such disturbed state, and
found that the key factor is the asymmetry factor $\alpha$ (described
below), which in fact can be calculated free from the model fitting,
though the center has to be determined by the model-fitting.

\subsubsection{Two key morphology parameters derived from Chandra X-ray images}

By looking at hundreds of X-ray images of clusters and the residual
images, we find that relaxed clusters generally have steeper radial
distribution of brightness than disturbed ones. Further more, relaxed
clusters generally have roundish morphologies while merging clusters
could show more elongated shapes. Thus, we define a new profile
parameter $\kappa$ for the brightness distribution as being
\begin{equation}
\kappa=\frac{1+\epsilon}{\beta}.
\label{kappa}
\end{equation}
Here, $\beta$ and $\epsilon$ are the power law index and the
ellipticity for the fitted $\beta$-model.

Disturbed clusters usually are more significantly asymmetric than
relaxed ones \citep[e.g.,][]{wod88,ozf+10,zof+10,rme13,wh13}. Here we
use the asymmetry factor $\alpha$ to reveal the dynamical state of
galaxy clusters, which is defined as being
\begin{equation}
\alpha=\frac{\sum\limits_{x_i,y_i}{[f_{\rm
        obs}(x_i,y_i)-f_{\rm
        obs}(x_i',y_i')]^2}}{\sum\limits_{x_i,y_i}{f^2_{\rm
      obs}(x_i,y_i)}} \times100\;  \%,
\label{alpha}
\end{equation}
where $f_{\rm obs}(x_i',y_i')$ stands for observed flux at the
symmetry pixel of $(x_i,y_i)$ with respect to the cluster center
$(x_0,y_0)$.

\begin{table*}
\begin{minipage}{1\linewidth}
  \label{tab2}
\centering
\caption{ Dynamical parameters for 125 clusters with known dynamical
  states from literatures (see
  http://zmtt.bao.ac.cn/galaxy\_clusters/dyXimages/ for the full table).}
\fontsize{12pt}{10} \renewcommand\tabcolsep{3pt}
\begin{spacing}{0.9}
\begin{tabular}{lrrrrcccccrc}
  \hline
 Name  & \mc{1}{c}{ObsID} & \mc{1}{c}{R.A.} & \mc{1}{c}{Dec.} & \mc{1}{c}{$z$}  & \mc{1}{c}{${\rm log}_{10}(c)$} &\mc{1}{c}{${\rm log_{10}}(\omega)$} &\mc{1}{c}{${\rm log_{10}}(P_3/P_0)$} &\mc{1}{c}{$\kappa$} &\mc{1}{c}{${\rm log}_{10}(\alpha)$} &\mc{1}{c}{$\delta$} & \mc{1}{c}{comment}\\
 \mc{1}{l}{(1)} &  \mc{1}{c}{(2)} &  \mc{1}{c}{(3)} &  \mc{1}{c}{(4)} &  \mc{1}{c}{(5)} &\mc{1}{c}{(6)} &  \mc{1}{c}{(7)} &  \mc{1}{c}{(8)} &  \mc{1}{c}{(9)} &  \mc{1}{c}{(10)} &  \mc{1}{c}{(11)} &  \mc{1}{c}{(12)}\\
\hline
MACS0011.7-1523      &  6105 &   2.9288  & -15.3894  & 0.3780  & -0.41$\pm$0.01 & -3.05$\pm$0.01 & -7.11$\pm$0.05 & 0.96 & -1.72$\pm$0.01 & -0.26$\pm$0.01 & R, 1 \\
A2744                &  8477 &   3.5883  & -30.3969  & 0.3014  & -0.99$\pm$0.01 & -1.68$\pm$0.01 & -6.11$\pm$0.03 & 1.57 & -0.58$\pm$0.01 &  0.96$\pm$0.01 & D, 4 \\
CL0016+1626          &   520 &   4.6408  &  16.4381  & 0.5410  & -0.80$\pm$0.01 & -2.39$\pm$0.04 & -6.88$\pm$0.10 & 1.53 & -1.25$\pm$0.01 &  0.48$\pm$0.01 & D, 3 \\
MACSJ0025.4-1222     & 10413 &   6.3725  & -12.3769  & 0.5843  & -0.82$\pm$0.01 & -2.02$\pm$0.02 & -6.34$\pm$0.05 & 1.97 & -1.26$\pm$0.01 &  0.79$\pm$0.01 & D, 4 \\
RXJ0027.6+2616*      & 14012 &   6.9575  &  26.2739  & 0.3668  & -0.74$\pm$0.01 & -2.17$\pm$0.01 & -6.66$\pm$0.04 & 1.74 & -1.11$\pm$0.01 &  0.73$\pm$0.01 & D, 3 \\
MACSJ0035.4-2015     &  3262 &   8.8608  & -20.2628  & 0.3640  & -0.61$\pm$0.01 & -2.14$\pm$0.01 & -7.43$\pm$0.08 & 1.13 & -1.75$\pm$0.01 & -0.16$\pm$0.01 & D, 3 \\
A68*                 &  3250 &   9.2785  &   9.1567  & 0.2537  & -0.69$\pm$0.01 & -1.81$\pm$0.01 & -6.61$\pm$0.02 & 1.57 & -1.37$\pm$0.01 &  0.42$\pm$0.01 & D, 2 \\
A2813                &  9409 &  10.8517  & -20.6214  & 0.2924  & -0.69$\pm$0.01 & -1.93$\pm$0.01 & -7.43$\pm$0.11 & 1.34 & -1.62$\pm$0.01 &  0.09$\pm$0.01 & D, 4 \\
Z348                 & 10465 &  16.7105  &   1.0697  & 0.2514  & -0.19$\pm$0.01 & -3.84$\pm$0.01 & -7.51$\pm$0.02 & 0.65 & -1.50$\pm$0.01 & -0.34$\pm$0.01 & R, 1 \\
MACSJ0111.5+0855     &  3256 &  17.8813  &   8.9275  & 0.2630  & -0.26$\pm$0.02 & -3.49$\pm$0.05 & -6.13$\pm$0.04 & 0.72 & -1.62$\pm$0.01 & -0.37$\pm$0.01 & D, 2 \\
\hline
\end{tabular}
{\\Notes: Columns: (1) cluster name; (2) observation ID of Chandra
  (3-4) right ascension and declination (J2000); (5) redshift; (6) the
  concentration index; (7) the centroid shift; (8) the power ratio;
  (9) the profile parameter; (10) the asymetry factor; (11) the
  morphology index; (12) Comments on dynamical state of clusters:
  ``R/D'' means relaxed or disturbed clusters, the number are
  dynamical flag classified by \citet{me12}. Clusters marked with
  ``*'' are selected from the BCS and eBCS catalogue
  \citep{eeb+98,eea+00} with the same selection criteria used by
  \citet{me12}, and dynamically categorized by us with the standard in
  \citet{me12}.}
\end{spacing}
\end{minipage}
\end{table*}

\begin{figure*}
  \begin{center}
\includegraphics[angle=0,width=0.75\textwidth]{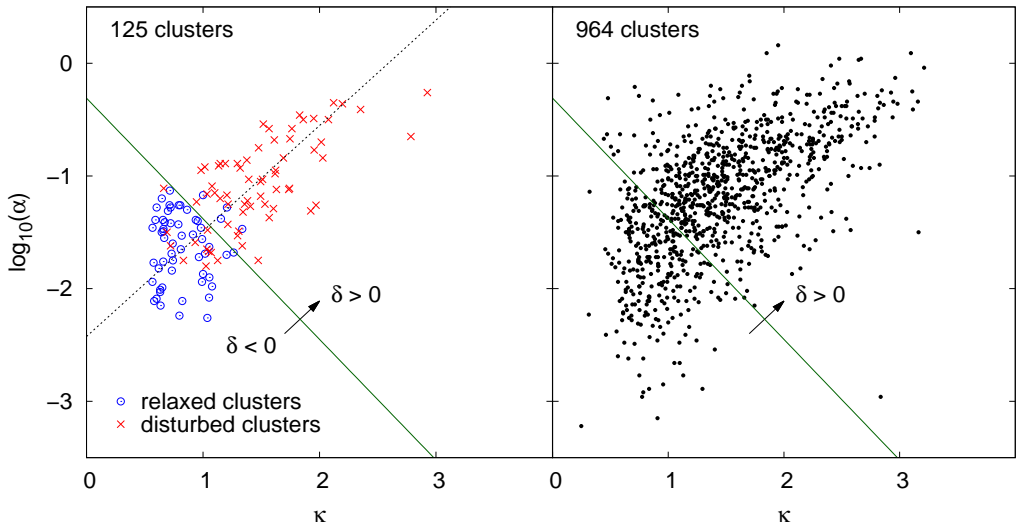}\\
\caption{{\it Left panel:} Distribution for a test sample of 125
  clusters in the $\kappa-\alpha$ space. The relaxed clusters are
  presented as circles while disturbed clusters as crosses. The dash
  line stands for the best fitted line for all sources in the test
  sample, the green solid line is the adopted boundary that separates
  relaxed and disturbed clusters with highest success rate. {\it Right
    panel:} Similar to the left panel but for 964 clusters.}
\label{ka}
\end{center}
\end{figure*}

\label{combpara}
Because the dynamical state cannot be represented with a single
parameter and an easy threshold (see Figure~\ref{ka}), the combination
of the profile parameter $\kappa$ and the asymmetry factor $\alpha$
can indicate the dynamical state of galaxy clusters more
properly. Here we define the morphology index $\delta$ as being:
\begin{equation}
\delta=A\kappa+B\alpha+C.
\label{delta}
\end{equation}

To find the appropriate values of the coefficients $A$, $B$ and $C$,
we take a test sample of 125 clusters with known dynamical states
qualitatively classified as relaxed or disturbed by \citet{me12} who
built a statistically complete sample from their MAssive Cluster
Survey \citep[MACS:][]{eeh01,ebd+07,eem+10} and the Brightest Cluster
Sample \citep[BCS \& eBCS:][]{eeb+98,eea+00} based on data of the {\it
  ROSAT All-sky Survey} \citep[{\it RASS}:][]{t93}, with a flux larger
than $1\times10^{-12}\rm erg ~s^{-1}cm^{-2}$ in 0.1-2.4 keV and a
luminosity $L_{\rm RASS}>5\times10^{44}\rm erg~s^{-1}$. They collected
129 clusters and got X-ray and optical images for 108 clusters of
them, and dynamically classified these clusters into 4 groups. We
directly take the dynamical information of the 108 clusters from
\citet{me12} and define clusters in group 1 as relaxed clusters and
those in group 2-4 as disturbed clusters. In addition, we take 17
extra clusters from the BCS and eBCS sample \citep{eeb+98,eea+00} that
satisfy $f_{\rm RASS}>1\times10^{-12}\rm erg ~s^{-1}cm^{-2}$ and
$L_{\rm RASS}>5\times10^{44}\rm erg~s^{-1}$ and have been observed by
{\it Chandra} later, and dynamically classify them with the same
standard in \citet{me12}. In total, we have 125 clusters with known
dynamical state as listed in Table 2. This test sample is X-ray flux
complete and volume complete to $z\sim0.4$ \citep{me12}.

\begin{figure*}
\begin{center}
\includegraphics[angle=-90,width=0.45\textwidth]{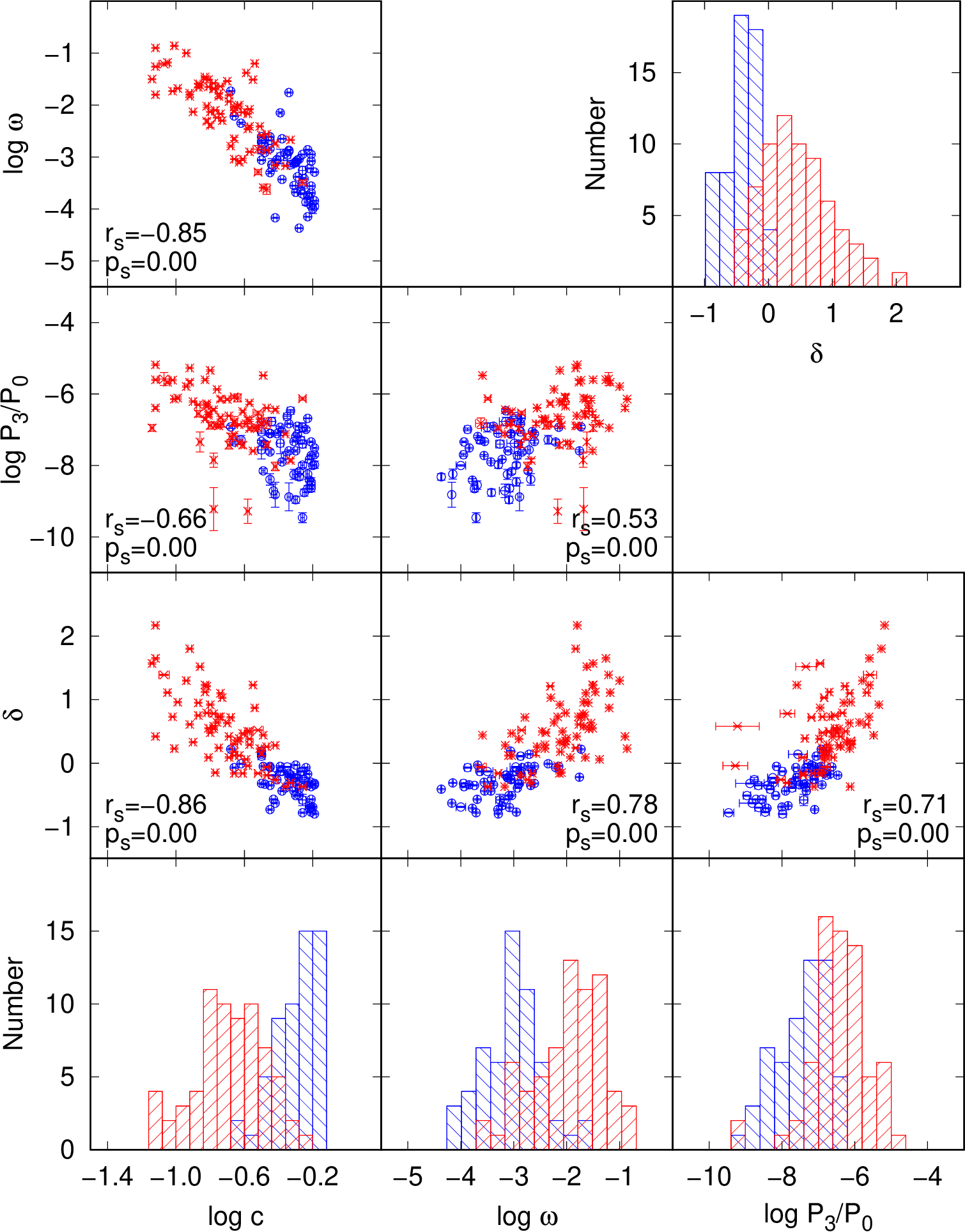}
\hspace{.4in}
\includegraphics[angle=-90,width=0.45\textwidth]{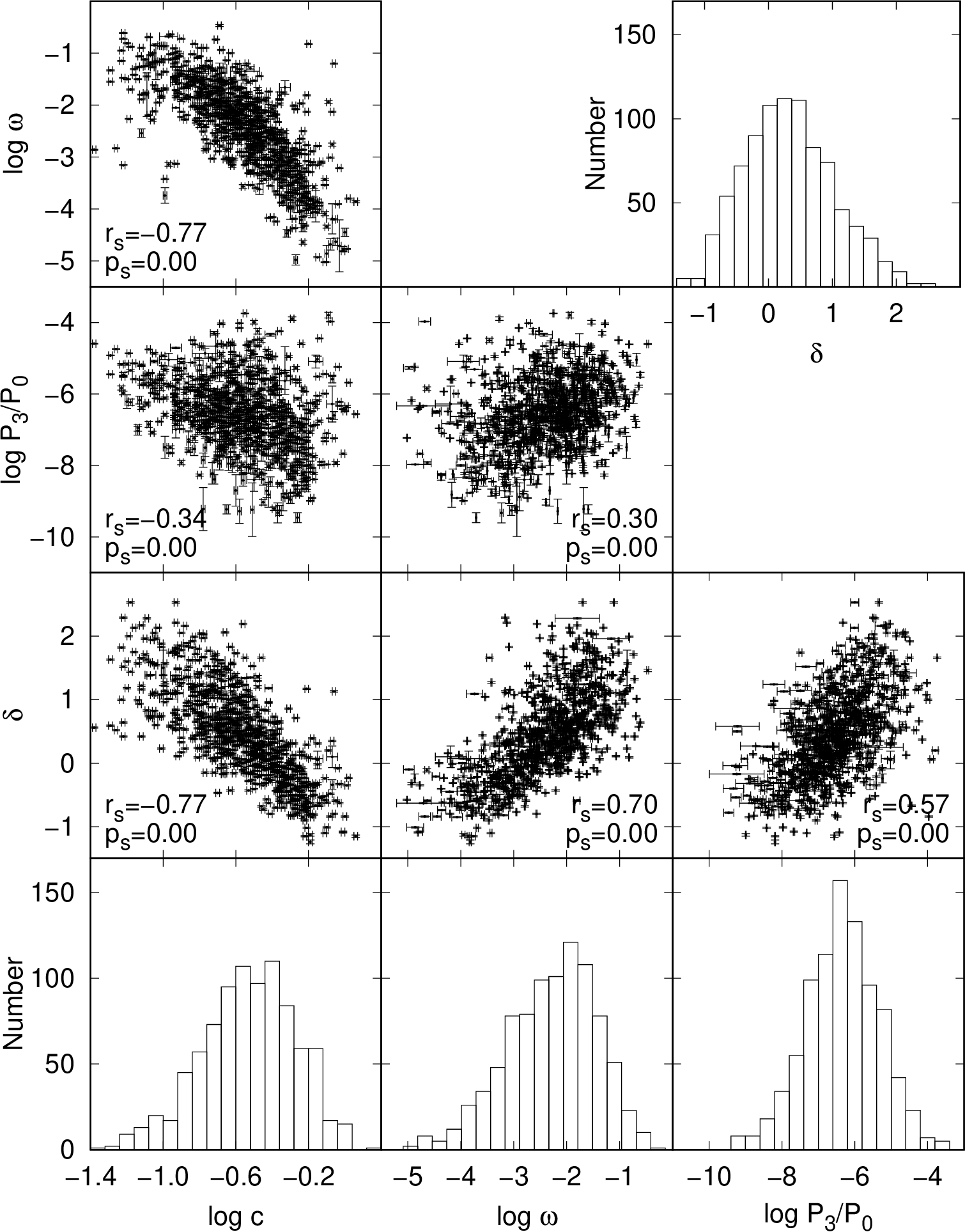}
\caption{{\it Left panel:} One-dimensional histogram distributions and
  two-dimensional correlations of four kinds of dynamical proxies for
  125 clusters with known dynamical states from literatures. Crosses
  and red histograms indicate disturbed clusters, while circles and
  blue histograms stand for relaxed clusters. The Spearman rank-order
  correlation coefficient $r_s$ and the relevant significance level of
  the correlation $p_s$ are labelled in each panel. A small value of
  $p_s$ indicates a significant correlation ($r_s$>0) or
  anti-correlation ($r_s$<0). {\it Right panel:} Similar to the left
  panel but for the sample of 964 clusters.
\label{corrsct}}
\end{center}
\end{figure*}

The $\kappa-\alpha$ distribution for these 125 clusters are plotted in
the left panel of Figure~\ref{ka}. Relaxed clusters are separated from
disturbed ones. First, in the $\kappa-\alpha$ space we find the
best-fitted line as
\begin{equation}
\rm{log_{10}}(\alpha)=0.94\kappa-2.43.
\label{kafit}
\end{equation}
The line of demarcation has the form of
\begin{equation}
{\rm log_{10}}\alpha=A_1\kappa+B_1,
\label{demarc}
\end{equation}
which is perpendicular to the best-fitted line with
$A_1=-1/0.94=-1.07$ and $B_1=-0.31$, and has the highest success rate
to discriminate the relaxed and disturbed clusters. For the test
sample the success rate is 115/125=88\%. The morphology index $\delta$
of galaxy clusters is therefore defined as the distance to the best
demarcation line in $\kappa-\alpha$ space, that is
\begin{equation}
\begin{aligned}
\delta &=0.68\rm{log_{10}}(\alpha)+0.73\kappa+0.21,
\label{delta}
\end{aligned}
\end{equation}
which can quantitatively indicate the dynamical state of galaxy
clusters just from the X-ray image without the redshift
information. We calculate the two parameters $\kappa$ and $\alpha$ and
hence the morphology index $\delta$ for all 964 clusters, as listed in
Table 1.

\section{Comparison and applications of the dynamical parameters}

\begin{figure}
\begin{center}
\includegraphics[angle=0,width=0.38\textwidth]{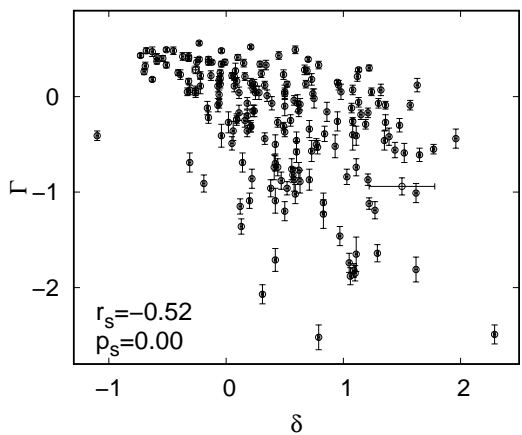}
\caption{Correlations between the X-ray morphology index, $\delta$,
  which we obtained in this paper and optical dynamical parameter,
  $\Gamma$, obtained by \citet{wh13} for 190 clusters. The Spearman
  rank-order correlation coefficient $r_s$ and the correlation
  significance $p_s$ are marked in the bottom left corner.}
\label{gcorr}
\end{center}
\end{figure}

\begin{figure*}
\begin{center}
\includegraphics[angle=0,width=0.65\textwidth]{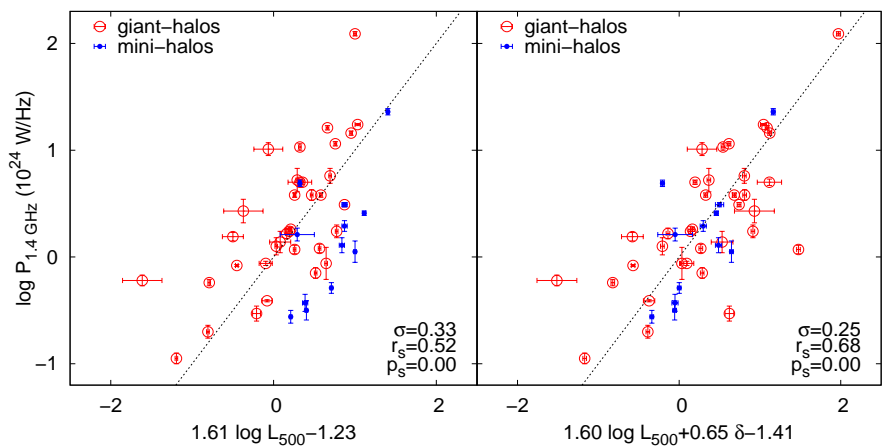}
\caption{Applications of the morphology index $\delta$ to the scaling
  relations for 35 clusters with giant-halos (open circles) and 12
  clusters with mini-halos (solid points).  The correlation is
  enhanced and the scatter is reduced when the dynamical parameter
  $\delta$ is involved ({\it right panel}) compared to the case
  without involvement ({\it left panel}). The dashed line stands for
  the best fitted line. The intrinsic scatter $\sigma$, the Spearman
  rank-order correlation coefficient $r_s$ and the correlation
  significance $p_s$ are marked in the bottom right corner.}
\label{halo}
\end{center}
\end{figure*}

\subsection{Comparison for the four dynamical parameters}
\label{paracorr}
Here we discuss the correlations among the four X-ray dynamical
parameters, the concentration index, the centroid shift, the power
ratio and the morphology index. To reflect the goodness of these
parameters more clearly, the test sample and the whole sample of
clusters are plotted and examined separately in Figure~\ref{corrsct}.

The left panels of Figure~\ref{corrsct} show that relaxed (blue) and
disturbed (red) clusters are separated in four parameters, indicating
that these four parameters are sensitive diagnostics for dynamical
state. However, as shown in the histograms they all are continuously
distributed without a clear boundary for the two states, which is
reasonable in practice. If one has to set a criterion to separate them
and get the fraction of disturbed clusters, that is 48.8\% if the
criterion is taken as $\delta>0$, close to the value of $\sim$50\%
obtained by previous works based on X-ray data
\citep[][]{bfs+05,crb+07,hmr+10}. In 2D parameter diagrams, clear
correlations are shown between each pair of two dynamical proxies,
though some of them have large scatter. The small $p_{s}$ in each
panel means the correlations are significant. In the right panels,
these dynamical parameters for the whole sample of 964 clusters are
found to be well correlated, and also distributed continuously from
very disturbed state to the relaxed state with a peak in
between. Strongest correlations appear between ${\rm log}_{10}(c)$,
$\delta$ and ${\rm log}_{10}(\omega)$, which implies that the three
parameters are similarly sensitive to the dynamical state.

\subsection{Correlation between the morphology index derived from X-ray image and the relaxation factor derived from the optical data}

Based on the projected distribution of member galaxies, \citet{wh13}
derived dynamical parameters for 2092 rich clusters. The optical image
of clusters is smoothed to 20 kpc with a Gaussian kernel and weighted
by luminosities of member galaxies. They got the relaxation factor
$\Gamma$ by combining the asymmetry factor, the ridge flatness and the
normalized deviation based on a test cluster sample with known
dynamical states from literatures. We get 190 clusters by cross
matching the optical catalogue in \citet{wh13} with our X-ray
sample. The Figure \ref{gcorr} shows the distribution of the optical
relaxation factor $\Gamma$ and the morphology index $\delta$. It is
clear that relaxed clusters are both well recognized in the view of
optical and X-ray dynamical parameters as the data are concentrated in
the upper-left corner. However, parameters are very scattered for
disturbed clusters, which means that both hot gas and member galaxies
can be very disordered but are not necessary to move or
distribute coincidentally.

\begin{figure*}
\begin{center}
\includegraphics[angle=-90,width=0.95\textwidth]{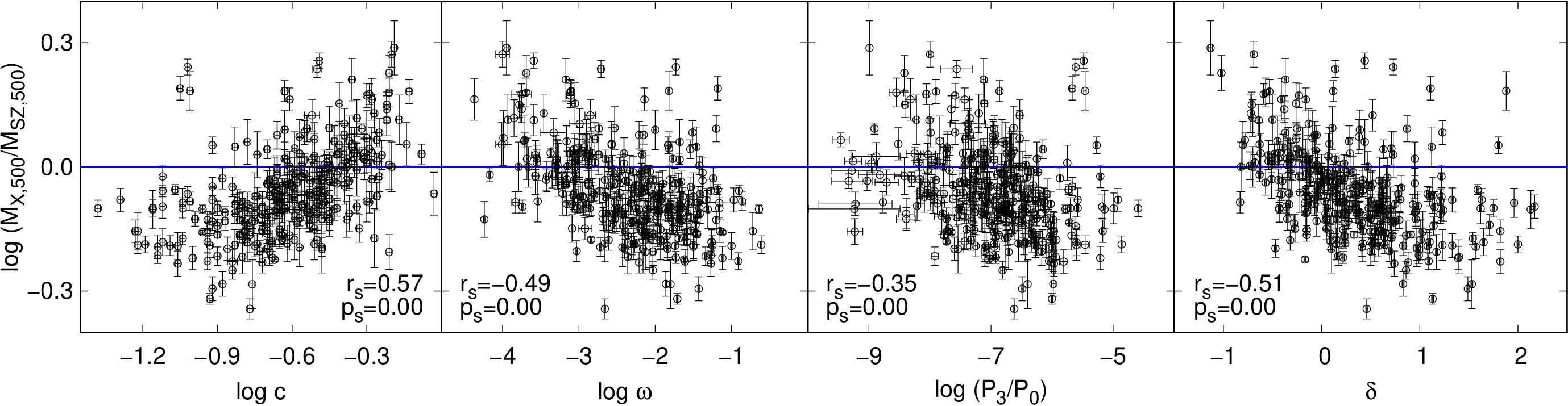}\\[3mm]
\includegraphics[angle=-90,width=0.98\textwidth]{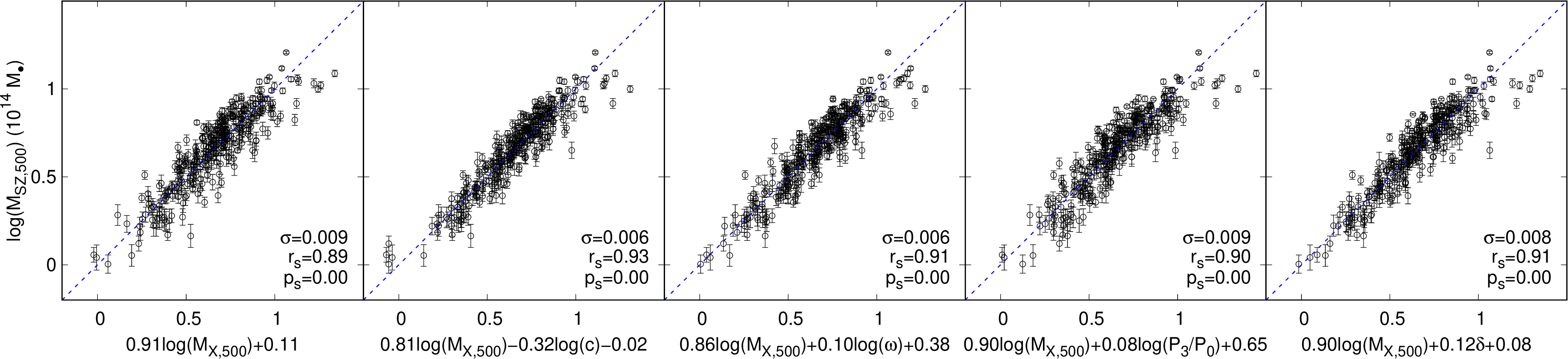}\\[3mm]
\includegraphics[angle=-90,width=0.98\textwidth]{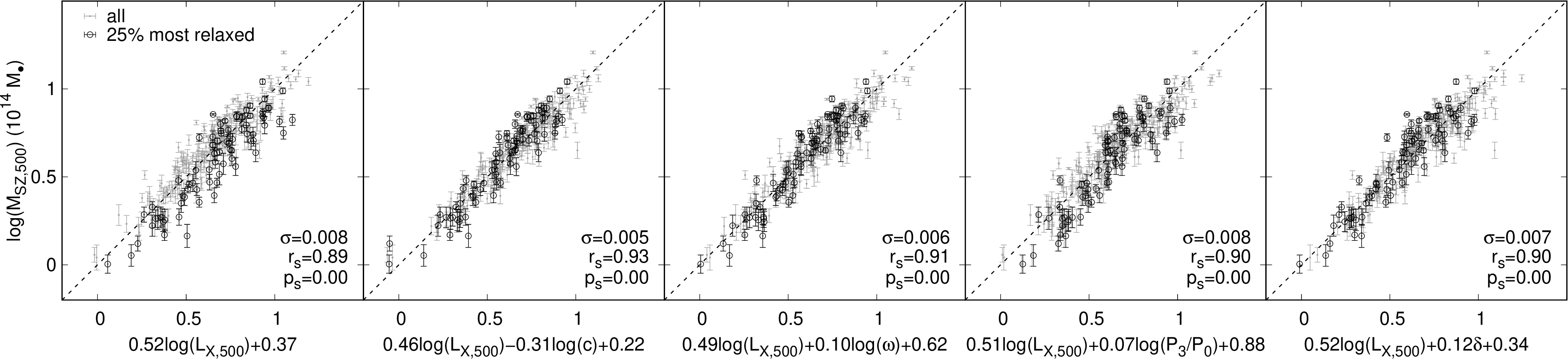}\\
\caption{{\it Top panels:} Relations between dynamical parameters and
  ratio of mass estimated from X-ray and SZ-effect data for 316
  clusters. The solid line is for the equality of the two estimated
  masses. {\it Middle panels:} Influence of dynamical parameters on
  the $M_{\rm X,500}-M_{\rm SZ,500}$ scaling relations for 316
  clusters. The dashed line is the best fitted line, and the intrinsic
  scatter $\sigma$, the Spearman rank-order correlation coefficient
  $r_s$ and the correlation significance $p_s$ are marked in the
  bottom right corner. {\it Bottom panels:} Similar to the middle
  panels but for the $L_{\rm X,500}-M_{\rm SZ,500}$ relation.  The
  25\% most relaxed clusters ($\delta<-0.15$) are denoted as black
  circles.}
\label{Moff}
\end{center}
\end{figure*}

\subsection{The scaling relations and the fundamental plane for radio giant-halos and mini-halos in clusters}

\begin{figure*}
\begin{center}
\includegraphics[angle=-90,width=0.95\textwidth]{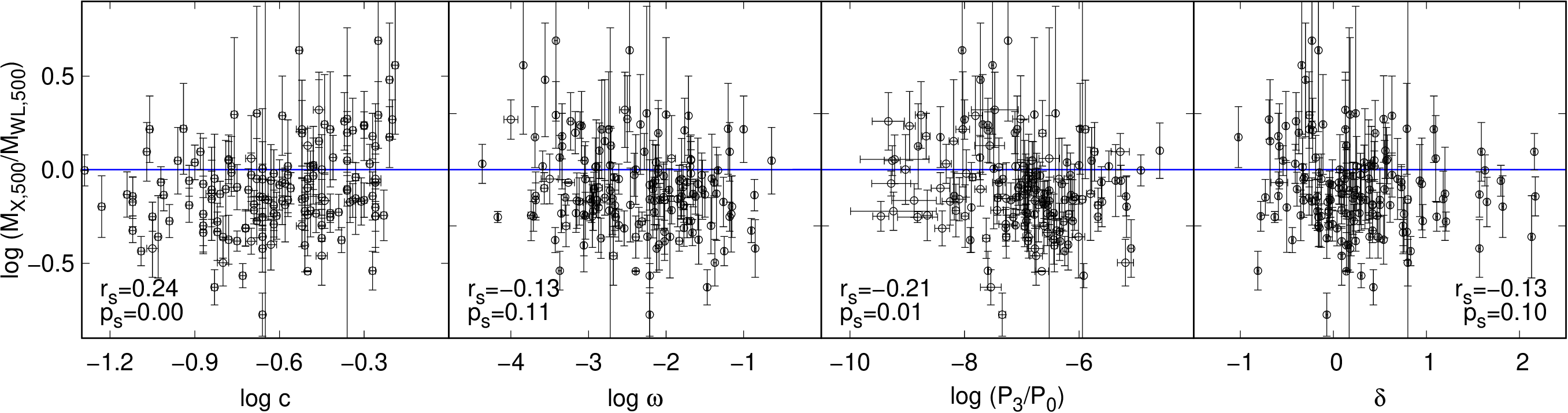}\\[3mm]
\includegraphics[angle=-90,width=0.98\textwidth]{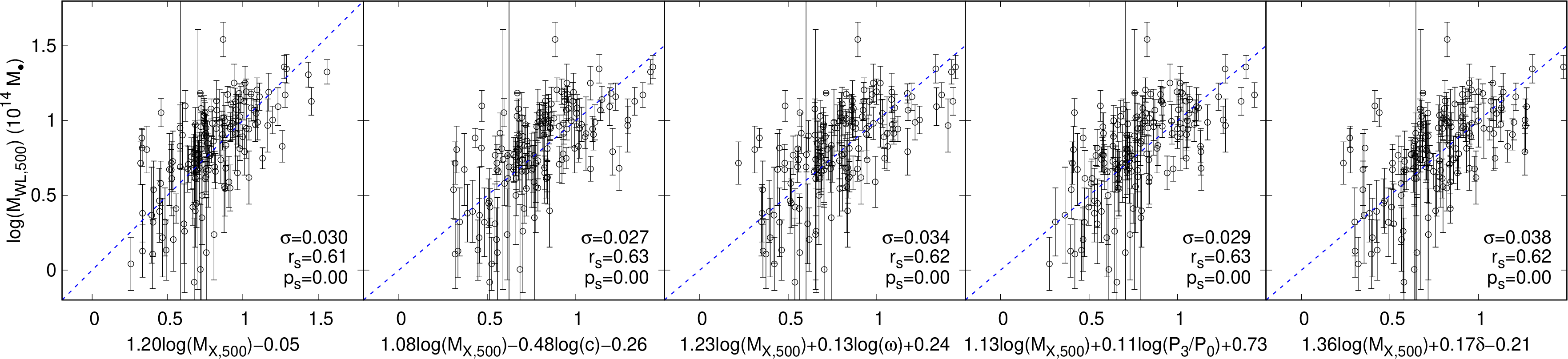}\\
\caption{Similar to Figure~\ref{Moff} but for masses estimated through
  weak lensing analyses.}
\label{Moff-wl}
\end{center}
\end{figure*}

Large scale diffuse radio sources in clusters are associated with ICM
rather than member galaxies, which can be classified into giant-halos,
relics and mini-halos \citep[see][as an observational
  review]{fgg+12}. It has been widely believed that the formation of
diffuse radio sources are related to merging process of their host
clusters. Giant-halos and relics are usually discovered in disturbed
clusters, while mini-halos are detected around the cool core of
relaxed clusters. Giant-halos and mini-halos may related to the
(re-)acceleration process by turbulence in the ICM, while relics are
related to the shock resulted by major merger \citep[see][for
  detail]{bj14}. The scaling relation has been established between
radio power of giant-halos and mini-halos and the mass or mass proxies
of the host clusters \citep[e.g.,][]{lhb+00,bcds09,ceb+13,yhw15}.

We collected the radio powers $P_{1.4~\rm GHz}$ in
$10^{24}\rm~W~Hz^{-1}$, the X-ray luminosities $L_{500}$ within
$R_{500}$ in $10^{44}\rm~erg~s^{-1}$ from \citet[][see the references
  therein]{yhw15}. Now we check if the morphology index $\delta$ can
reduce the scatter of data and enhance the correlation for the scaling
relation. In the left panel of Figure~\ref{halo}, the radio power of
giant-halos (red open circles) for 35 clusters and mini-halos (blue
solid points) for 12 clusters are plotted against X-ray luminosity of
host clusters. The mini-halos have statistically less radio power than
giant-halos but follow the same scaling relation with the X-ray
luminosity. We fit all these data with weight of data errors, see the
Appendix in \citet{yhw15}. The scaling relation has a Spearman
rank-order correlation coefficient $r_s=0.52$. In the right panel, the
morphology index is involved and the scaling relation show a larger
correlation index ($r_{\rm s}=0.68$) and a reduced scatter
($\sigma=0.25$), which means that giant-halos and mini-halos follow
the similar fundamental plane between the radio power, cluster mass
and the morphology index.

\subsection{Influence of dynamical state on mass estimation for clusters}

The gravitational mass of clusters is an essential parameter for a lot
of researches, which is however affected by the dynamical state if
they are estimated from X-ray images. \citet{pap+11} published a
meta-catalogue contains 1,743 clusters based on previous works from
data of {\it ROSAT} All Sky Survey. The masses $M_{\rm X,500}$ and
characteristic radius $R_{500}$ of clusters are estimated
homogeneously by using X-ray data. \citet{paa+16} detected 1,653
clusters and estimated masses $M_{\rm SZ, 500}$ from SZ-effect maps,
which is insensitive to dynamical states of clusters
\citep[e.g.,][]{mhbn05}. Here we use a sample of 316 clusters commonly
in the catalogues in \citet{pap+11} and \citet{paa+16} to check the
influence of dynamical state on the mass estimation for clusters.

In the top panels of Figure~\ref{Moff}, we show the distribution of
four dynamical parameters against the ratio of masses estimated from
X-ray and SZ-effect data, and find a clear dependence on the
concentration index $c$, the centroid shift $\omega$ and the
morphology index $\delta$, though the correlation is weaker for the
power ratio $P_3/P_0$. The mass estimated from the X-ray images are
underestimated comparing to masses derived from SZ-effect for
disturbed clusters, but apparently overestimated for relaxed
clusters. The SZ-effect reflects the thermal and non-thermal
components of clusters simultaneously. The underestimation of mass
through X-ray data for disturbed clusters may indicate that disturbed
clusters have higher fraction of energy stored as non-thermal form. In
the middle rank of Figure~\ref{Moff}, the influence of dynamical
parameters is considered for the correction of the mass estimation
from the X-ray images on the $M_{\rm X,500}-M_{\rm SZ,500}$
relation. The Spearman rank-order correlation coefficient increase
when dynamical parameters are considered.

On the other hand, the X-ray luminosity $L_{\rm X,500}$ is widely used
as a mass proxy of clusters. In the bottom panels of
Figure~\ref{Moff}, we show the $L_{\rm X,500}-M_{\rm SZ,500}$
relations for the 316 clusters. Since relaxed and disturbed clusters
usually show different mass scaling relations
\citep[e.g.,][]{omb+06,ldc+06,ldk+09,zjc+13}, we compare the 25\% most
relaxed clusters ($\delta<-0.15$, black circles) with the full
sample. For the 2D correlation (the most left panel), the relaxed
clusters are clearly offset from the best fitted line determined by
the total sample. The offset is significantly reduced when the
dynamical parameters included (the right four panels), which again
means that there is a fundamental plane for the mass, luminosity and
dynamical state for galaxy clusters. The 3D correlations have smaller
intrinsic scatter and larger $r_{\rm s}$ than the 2D correlation.

For the further demonstration, we investigate the influence of
dynamical state to the lensing mass of galaxy clusters which is
believed to be the most reliable. \citet{s15} compiled a catalogue
which contains 485 clusters with mass ($M_{\rm WL,500}$) estimated
through weak lensing analyses. We cross-match our samples with the
catalogues in \citet{s15} and \citet{pap+11}, and get 151 clusters in
common. In Figure~\ref{Moff-wl}, we do similar correlations to
Figure~\ref{Moff} but replace $M_{\rm X,500}$ with $M_{\rm
  WL,500}$. In the upper panels of Figure~\ref{Moff-wl}, the mass
ratio ($M_{\rm X,500}/M_{\rm WL,500}$) is not randomly around but
mostly less than 0, which means that the X-ray mass is
underestimated. In the lower panels, the correlations are enhanced
with slightly larger $r_s$ when dynamical parameters are included.

\section{Summary}
The {\it Chandra} satellite has accumulated X-ray image data for
$\sim$1000 clusters of galaxies, which make it feasible to work out a
catalogue of dynamical parameters for a large sample of clusters.

We collected Chandra data for 964 galaxy clusters. The images for
these clusters are processed following the same procedure, and
smoothed by a Gaussian function with a size of 30 kpc (or 5 arcseconds
for few clusters with no redshift available). Three widely used
dynamical parameters, i.e., the concentration index $c$, the centroid
shift $\omega$ and the power ratio $P_3/P_0$, are calculated from the
X-ray images in a circular region with a radius of 500 kpc. Our
results are consistent with the values available in literatures. We
also derive two adaptive parameters, the profile parameter $\kappa$
and the asymmetry factor $\alpha$ in the best fitted elliptical
region, and define the morphology index $\delta$, which can be
excellent indicator for dynamical state. Finally, we calculate the
profile parameter $\kappa$, the asymmetry parameter $\alpha$ and the
morphology index $\delta$ for 964 clusters (see Table 1).

The four dynamical parameters are well correlated, and can separate
relaxed clusters from disturbed ones. But they are distributed
continuously from the most disturbed status to the relaxed. The
morphology index we derived from the X-ray images and optical
relaxation factor is also significant correlated.

By involving the dynamical parameters, we find that clusters with
diffuse radio giant-halos and mini-halos follow the same fundamental
plane between radio power, cluster mass and morphology index. Mass
estimation for cluster of galaxies if derived from X-ray data is
affected by their dynamical states.

\section*{Acknowledgements}

We thank the referee and Dr. Z. L. Wen for helpful comments. The
authors are supported by the National Natural Science Foundation of
China (11803046, 11988101 and 11833009) and the Young Researcher Grant
of National Astronomical Observatories, Chinese Academy Sciences.
This research has made use of data obtained from the {\it
  Chandra Data Archive} and the {\it Chandra Source Catalog}, and
software provided by the {\it Chandra X-ray Center (CXC)} in the
application packages {\it CIAO}, {\it ChIPS}, and {\it Sherpa}.

\section*{Data availability}

The data underlying this article, the X-ray images for 964 clusters,
and also the code for calculations are available on the web-page
http://zmtt.bao.ac.cn/galaxy\_clusters/dyXimages/.

\bibliographystyle{mnras}
\bibliography{ref}
\label{lastpage}
\end{document}